# Formation of Iron-Helium Compounds under High Pressure

Haruki Takezawa,[1,*] Han Hsu,[2,†] Kei Hirose,[1,3,‡] Fumiya Sakai,[1] Suyu Fu,[1] Hitoshi Gomi,[3] Shiro Miwa,[4] and Naoya Sakamoto[4]

[1]*Department of Earth and Planetary Science, The University of Tokyo, Tokyo 113-0033, Japan*
[2]*Department of Physics, National Central University, Taoyuan City 320317, Taiwan*
[3]*Earth-Life Science Institute, Institute of Science Tokyo, Tokyo 152-8550, Japan*
[4]*Creative Research Institution (CRIS), Hokkaido University, Sapporo, Hokkaido 001-0021, Japan*

To whom correspondence should be addressed:
Haruki Takezawa (takezawa@eps.s.u-tokyo.ac.jp), Han Hsu (hanhsu@ncu.edu.tw), Kei Hirose (kei@eps.s.u-tokyo.ac.jp).

**Abstract**

We report the formations of fcc and distorted hcp iron-helium compounds with $x$ in FeHe$_x$ up to 0.13 and 0.48, respectively, based on experiments at 5–54 GPa and ~1000–2820 K. Upon releasing pressure under room temperature, these fcc and distorted hcp FeHe$_x$ were still observed by XRD and SIMS measurements. Our first-principles calculations indicate that fcc and hcp FeHe$_x$, with helium atoms occupying the tetrahedral and trigonal-planar interstitial sites (instead of the octahedral sites), are dynamically stable throughout 0–50 GPa. These results support that the Earth's core can be a large reservoir of primordial $^3$He.

**Main text**

*Introduction.*—Both theory and experiment have revealed that noble gases, in particular heavy ones, are no longer inert under high pressure by changing their electronegativity and reactivity, forming a variety of compounds including those with noble gas insertion, as reviewed by Refs. [1,2]. For instance, the formation of XeFe$_3$ was predicted [3] and experimentally confirmed above 200 GPa and 2000 K [4]. In addition, xenon-iron oxides were found to exist above 150 GPa by theory [5], and the incorporation of xenon into SiO$_2$ was observed by experiments at pressures as low as ~1 GPa [6,7]. On the other hand, helium compounds are little known. Na$_2$He was reported by calculation and experiment above 113 GPa [8], while FeO$_2$He (>120 GPa) [9] and MgF$_2$He (>107

GPa) [10] were theoretically predicted. Other known helium-bearing compounds are van der Waals (vdW) molecules: He(N$_2$)$_{11}$ [11], H$_2$O-He [12], NH$_3$-He [13,14], and CH$_4$-He [15].

Reactions of helium with iron under high pressure and temperature (*P-T*) have been investigated through the partitioning of helium between liquid iron and molten silicate by experiment [16,17] to 16 GPa/3000 K and by theory [18–21] to 135 GPa/5000 K. Even though these results consistently show the partitioning of a minor amount of helium into liquid iron, it has been argued that the Earth's core, composed mainly of iron, is an important reservoir of primordial $^3$He. To our knowledge, neither iron-helium compounds nor Fe lattice with He insertions have been experimentally explored at high pressures, while one theoretical prediction for iron-helium compounds at the terapascal pressure range has been reported [22].

In this study, iron–helium reactions at high *P-T* were examined in a laser-heated diamond-anvil cell (DAC), based on synchrotron x-ray diffraction (XRD) measurements at beamline BL10XU, SPring-8. We observed the appearance of the face-centered cubic (fcc) and (distorted) hexagonal close-packed (hcp) phases whose unit-cell volumes were remarkably larger compared to pure Fe at equivalent pressure. Such volume expansion is attributed to the incorporation of helium atoms into the interstitial sites in the fcc and hcp lattices of iron and the possible formation of FeHe$_x$ compounds ($x$ = 0.05 to 0.48). When the pressure was released, the expanded fcc and (distorted) hcp phases persisted, while the helium contents in the former were lower than observed at high pressures. As a complement to these XRD measurements, we carried out secondary ion mass spectrometry (SIMS) analyses of helium in a recovered Fe-He sample under cryogenic temperatures and found its concentration comparable to the estimate based on XRD. To further investigate FeHe$_x$ compounds, we performed first-principles calculations based on density functional theory (DFT). Our calculations indicate that fcc and hcp FeHe$_x$, with He atoms occupying the tetrahedral (T) and trigonal-planar (F) interstitial sites (instead of the octahedral (O) sites), are dynamically stable in various magnetic states throughout 0–50 GPa and are expanded in volume, in support of the experimental observations.

*Experimental results.*—We carried out five separate experiments at 5–54 GPa and ~1000–2820 K (Table I, Table SI [23], see High *P-T* experiments [23]). In the first run with only Fe foil and He gas originally in a sample chamber (Fig. S1 [23]), when the sample was compressed to 15 GPa, the XRD pattern showed the coexistence of body-centered cubic (bcc) iron and solid hcp helium [Fig. 1(a)]. Subsequently we heated the sample to 1450 K for 10 min and observed the appearance of the volume-expanded hcp phase [Fig. 1(b)]. It exhibited a splitting of the hcp 100 and 101 peaks upon quenching temperature [Fig. 1(c)], likely because of slight crystallographic distortion to an orthorhombic structure [24]. The unit-cell volume of this distorted hcp phase was larger by 30% than that of pure Fe at equivalent pressure (Fig. 2). Considering such volume expansion was caused by the incorporation of helium into the interstitial sites around Fe atoms (see *Theoretical calculations* section below), the amount of helium in FeHe$_x$ is calculated to be $x$ = 0.48 based on $\Delta V_{He}$ (volume increase of Fe per He atom) (Fig. S2 [23]) for the ferromagnetic (FM) state (see Table SI and High *P-T* experiments [23]).

In run #2, we compressed an Fe foil with He to 21 GPa (Fig. S3 [23]) and then melted them by heating to 2820 K, which is higher than the melting point of pure Fe [25,26]. The XRD data showed diffuse scattering signal (Fig. S4 [23]), characteristic of liquid, along with the weak peaks from the fcc phase. We estimated the 2π/Q_distance from the first diffuse ring (the first peak position in structure factor S(Q) given in Fig. S5a [23]), which is relevant to the Fe-Fe distance in liquid, and found that it is longer than that for liquid Fe at equivalent pressure [27,28] (Fig. S5b [23]), indicating the elongation of the Fe-Fe distance by incorporating helium into the interstitial site. After quenching the sample to 300 K at 22 GPa, we observed new fcc and hcp phases (Fig. S3 [23]), both of which formed from liquid as "quench crystals". The unit-cell volumes of these fcc and hcp quench crystals were expanded compared to those of corresponding pure Fe phases (Fig. 2), suggesting that liquid iron also incorporates helium. Their helium contents are $x$ = 0.13 and 0.28 in FeHe$_x$, respectively (helium concentration in liquid may be between the two), considering the antiferromagnetic (AFM) state for fcc and the FM state for hcp, which are higher than that of fcc FeHe$_{0.05}$ coexisting with liquid during heating (Table SI [23]). This sample was recovered at ambient condition and examined by SIMS under cryogenic temperatures (see SIMS and EPMA analyses [23]). The SIMS measurements showed about 1.0 wt% He in the maximum ~1.5 μm thick sample (blue line in Fig. S6 [23]), consistent with the estimate of helium concentration by XRD between $x$ = 0.13 and 0.28 (or 0.9 and 2.0 wt%) in liquid. We also note that such cryogenic analysis contrast that performed under non-cryogenic conditions. In the latter, we found a substantial amount of helium at the surface of a recovered sample but negligible concentrations in its inside (black line in Fig. S6 [23]), indicating that helium escaped from the Fe lattice and accumulated at the sample surface during ion beam irradiation under non-cryogenic temperatures.

We decompressed the sample to ambient pressure with XRD measurements after forming the volume-expanded fcc and/or distorted hcp phases in runs #3 and #4. In the former experiment, we heated the Fe + He sample to ~1000 K at relatively low pressure of ~5.3 GPa and found that bcc Fe fully transformed into distorted hcp FeHe$_{0.33}$ in the FM state (Fig. S7 [23]). Upon complete pressure release, the distorted hcp FeHe$_{0.32}$ was still observed, exhibiting a lattice volume remarkably larger than that of hcp pure Fe (Fig. 2). Its helium content practically did not change during decompression (Table I). In run #4, both fcc and distorted hcp FeHe$_x$ were recovered at 1 bar (Fig. S8 [23]). Contrary to distorted hcp, fcc exhibited a partial loss of helium from $x = 0.09$ at 12–19 GPa to $x = 0.06$ at 1 bar (Table I). A small amount of the bcc phase appeared upon pressure release, exhibiting the unit-cell volume equivalent to that of bcc pure Fe and thus indicating the absence of helium in it. Additionally, fcc FeHe$_{0.06}$ was formed at 54 GPa in run #5, the highest pressure in the present experiments (Fig. S9 [23]). On a cross section of the recovered sample, we found minimal contamination by carbon (diamond anvils), oxygen, rhenium (gasket), and tungsten (needle used for loading an iron foil) (Fig. S10 and Table SII [23]). See more descriptions of runs #4 and #5 in Supplementary Material.

*Theoretical calculations.*— In our DFT calculations, fcc FeHe$_{0.25}$ and hcp FeHe$_{0.167}$ are considered (see Computational methods [23] and Refs. [29,30]). Remarkably, we find the site occupancy of helium completely different from hydrogen. In fcc/hcp FeH$_x$, H atoms predominantly occupy the O-site [31–33]. In contrast, fcc FeHe$_{0.25}$ with O-site He is dynamically unstable (Fig. S11 [23]). Further analysis indicates that for the soft phonon modes, atomic displacements of He are predominant, while Fe displacements are insignificant. At the Γ-point, the He displacements for the three degenerate soft phonon modes [Figs. S11(b)–S11(d) [23]], are along the [100], [010], and [001] directions, respectively. By displacing the He atom along the [111] direction followed by structural optimization, the He atom is subsequently equilibrated at the T-site [Figs. 3(a) and S12(a) [23]]. For fcc FeHe$_{0.25}$ with T-site He, various magnetic states are obtained (Fig. S12 [23]). Among them, the double-layer AFM (AFM-d) state [Fig. S12(b) [23]] is of interest, for it is the ground state of fcc Fe (Fig. S13 [23] and Refs. [34,35]). Relative enthalpies ($\Delta H$) of the FM, AFM-d, and nonmagnetic (NM) states are plotted [Fig. 3(c), NM as the reference], for these states are energetically favorable and dynamically stable in the relevant pressure region [Figs. 3(d)–3(f)]. Given that FM fcc Fe is dynamically unstable at 0 GPa [36], the FM ground state of fcc FeHe$_{0.25}$ [Fig. 3(c)] indicates that T-site He helps stabilize FM ordering in fcc Fe at low pressure. Above ~8 GPa, AFM-d FeHe$_{0.25}$ is energetically favorable. Despite significant tetragonal stretch in AFM-d fcc Fe [34,35], AFM-d FeHe$_{0.25}$ is nearly cubic [Fig. S12(f) [23]]. Upon compression, Fe magnetic moment ($\mu_{Fe}$) of AFM-d FeHe$_{0.25}$ decreases [Fig. S12(c) [23]], merging into the NM state at ~50 GPa [Fig. 3(c)]. Beyond ~50 GPa, NM FeHe$_{0.25}$ still remains dynamically stable.

In hcp Fe, helium exhibits complicated site occupancy. NM hcp FeHe$_{0.167}$ with O-site

He is dynamically stable at high pressure but unstable below 50 GPa [Figs. S14(a), (b) [23]]. The Γ-point soft phonon mode [Fig. S14(b) [23]] corresponds to He displacement along the [001] direction; by displacing the He atom accordingly, followed by structural optimization, the He atom is subsequently equilibrated at the F-site (center of a trigonal plane connecting two octahedra) [Figs. 3(g) and S14(c) [23]]. Such a structure is dynamically stable at low pressure (down to ~0 GPa) but unstable at high pressure [Figs. 3(k) and S14(d) [23]]. Remarkably, this high-pressure soft phonon mode also corresponds to He displacements along [001]. These results indicate that a pressure decrease from >100 to <50 GPa may induce He diffusion from the O-site to the F-site. Another possible structure, NM hcp $FeHe_{0.167}$ with T-site He, is dynamically stable [Figs. S14(e), (f) [23]] but energetically unfavorable [Figs. 3(i) and S15 [23]].

Below ~20 GPa, hcp Fe has nonzero $\mu_{Fe}$, but its magnetic ordering has been controversial [37–39]. Furthermore, in fcc Fe, He stabilizes FM ordering at low pressure [Fig. 3(c)]. Therefore, we simply consider FM ordering when testing the dynamical stability of magnetic hcp $FeHe_{0.167}$. At low pressure, FM hcp $FeHe_{0.167}$ with O-site or F-site He are both dynamically unstable [Figs. S16(a), (b) [23]]. For $FeHe_{0.167}$ with T-site He (*P31m*, Fig. S14(e) [23]), when $\mu_{Fe} \neq 0$, the three Fe2 atoms shift to a Wyckoff position with higher symmetry. The resultant *P6mm* structure [Fig. 3(h)] is dynamically stable and energetically favorable at <20 GPa [Figs. 3(i), (j)]. Given the above results, our calculations sufficiently support the experimental observations: fcc and hcp $FeHe_x$ can be quenched to room temperature throughout 0–50 GPa. The discrepancy between the observed and computed hcp $FeHe_x$ structures may result from different He concentrations.

To analyze the He-Fe bonding, electron localization functions (ELFs) [40–42] of $FeHe_x$ are plotted [Figs. 3(a), 3(g), and 3(h)] for the respective states/pressures indicated in Figs. 3(d), 3(f), 3(j), and 3(k). For comparison, the ELF of FM fcc Fe is also plotted [Fig. 3(b)], which exhibits characteristics of metallic bonds, including ELF ≈ 0.4 in the interstitial regions and the presence of non-nuclear maxima [42]. These features can also be observed on the He-free lattice planes in hcp $FeHe_{0.167}$, including $z = 0.5$ in the NM (F-site) and $z = 0.309$ in the FM (*P6mm*) states [Figs. 3(g), (h)]. In the regions between Fe and He atoms, ELF ≈ 0, indicating vdW interactions [42] (see Fig. S17 [23] for finer details). The induced dipole moments (for vdW interactions) are formed via charge redistribution and slight charge transfer, as can be observed from the distortion of the ELFs and the charge density difference $\rho_{diff} \equiv \rho_{crystal} - \rho_{atoms}$ (Fig. S18 [23], $\rho_{crystal}$ and $\rho_{atoms}$ are the electron densities of the crystal and superposition of isolated atoms, respectively).

*Formation of Fe-He compounds.*—The present experiments demonstrate that iron reacts with helium to form Fe-He compounds at pressures as low as ~5.3 GPa and ~1000 K. Their dynamical stability and He-Fe bonding have been confirmed by our DFT calculations. While it may be analogous to the xenon incorporation to hcp Fe above ~300 GPa [43] and the formation of $XeFe_3$ at >200 GPa [3,4], the Fe-He compounds form at much lower pressures.

The volumes of helium-incorporated fcc and (distorted) hcp phases are provided as a function of pressure in Fig. 2, demonstrating that the latter always exhibited a larger volume per Fe atom (indicating higher helium concentration) than the former at a given pressure. Furthermore, we obtained the coexistence of fcc $FeHe_{0.09}$ and helium-richer distorted hcp $FeHe_{0.33}$ in run #4. These observations might indicate that the fcc and (distorted) hcp phases appear at relatively He-poor and He-rich portions of the Fe-He phase diagram around 20 GPa, respectively. It contrasts the phase relations in the Fe-FeH system, where H-rich fcc coexists with H-poor hcp above ~30 GPa [44]. These results also indicate that the presence of helium expands the stability field of hcp with respect to fcc and bcc, which is consistent with the formation of (distorted) hcp Fe-He at ~5.3 GPa and ~1000 K, where the fcc phase is stable in pure Fe.

*Helium in iron cores of terrestrial planets.*—These results stand for the presence of helium in planetary metallic cores. The present experiments show the formation of both liquid and solid Fe-He compounds to 54 GPa in pressure and 2820 K in temperature. Since the observed fcc and (distorted) hcp structures are close-packed crystal structures, these solid Fe-He phases, along with its liquid form, are most likely stable to Earth's core conditions (>135 GPa, >~4000 K). During Earth's accretion, a large fraction of primordial $^3$He could have been sequestrated into the core via metal-silicate chemical equilibrium in a deep magma ocean under high *P-T* (e.g., Refs. [16,17,21]). Earlier experiments [16] argued very low metal/silicate partition coefficient of helium and found only up to 6 ppm He in quenched liquid Fe even when no silicate was present. However, the depth profile obtained by their laser-ablation mass spectrometry measurements [16] is similar to that found by the present SIMS analysis under non-cryogenic temperatures (black line in Fig. S6 [23]), suggesting that helium was lost from iron in both cases. It is therefore possible that the metal/silicate partition coefficient of helium is much higher than reported [16], supporting the recent argument that the high $^3$He/$^4$He ratios observed in OIBs (e.g., Ref. [45]) are attributed to the core as an important reservoir of primordial $^3$He, instead of undegassed materials in the deep lower mantle [46,47].

Synchrotron XRD measurements were carried out at BL10XU, SPring-8 (proposals no. 2023A0181, 2023B0306, and 2024A0306). We thank K. Oka, N. Ikuta, and S. Mita for valuable discussion and technical assistance. Comments from three anonymous referees were valuable to improve the manuscript. This work was supported by JSPS grant 21H04506 to K.H.; H.H. is supported by National Science and Technology Council of Taiwan under grants NSTC 113-2116-M-008-010, 112-2112-M-008-038, and 111-2112-M-008-032. Part of the calculations were performed on the facilities of National Center for High-performance Computing (NCHC) of Taiwan. All data supporting the results of this study are available from the corresponding author upon reasonable request.

TABLE I. Experimental results.

| | $P$ (GPa) by He | $T$ (K) | Observed phases | fcc $x$ | hcp, dis-hcp[c] $x$ |
|---|---|---|---|---|---|
| 1 | 15.5 | 1450 | hcp Fe-He, hcp Fe | | |
| | 14.6 | 300 | dis-hcp[c] Fe-He$_{0.48}$, hcp Fe, solid He | | 0.481 |
| 2 | 30.5 | 2820 | fcc Fe-He | | |
| | 21.7 | 300 | fcc(q)[d] FeHe$_{0.13}$, fcc FeHe$_{0.05}$, hcp(q)[d] FeHe$_{0.28}$, solid He | 0.127 0.052 | 0.283 |
| 3 | ~5.3[a] | ~1000 | dis-hcp[c] Fe-He | | |
| | 7.4[a] | 300 | dis-hcp[c] FeHe$_{0.33}$ | | 0.334 |
| | 0.0 | 300 | dis-hcp[c] FeHe$_{0.32}$ | | 0.318 |
| 4 | 23.7 | 1660 | fcc Fe-He, dis-hcp[c] Fe-He | | |
| | 18.5 | 300 | fcc FeHe$_{0.09}$, dis-hcp[c] FeHe$_{0.33}$, solid He | 0.091 | 0.326 |
| | 11.6[b] | 300 | fcc FeHe$_{0.09}$, dis-hcp[c] FeHe$_{0.32}$ | 0.089 | 0.319 |
| | 0.0 | 300 | fcc FeHe$_{0.06}$, dis-hcp[c] FeHe$_{0.28}$, bcc Fe | 0.063 | 0.276 |
| 5 | 53.5 | 2140 | fcc Fe-He, hcp Fe | | |
| | 37.9 | 300 | fcc FeHe$_{0.06}$, hcp Fe, solid He | 0.055 | |

[a]Pressure was determined from bcc Fe present in unheated portion. [b]Pressure obtained from Al$_2$O$_3$. [c]Distorted hcp (orthorhombic) structure with the axial ratio of $b/\sqrt{3}a$ = 1.012 to 1.031 while 1.0 for hcp, showing the distortion in the *a-b* plane (see Table SIII [23]). [d]Quench crystals.

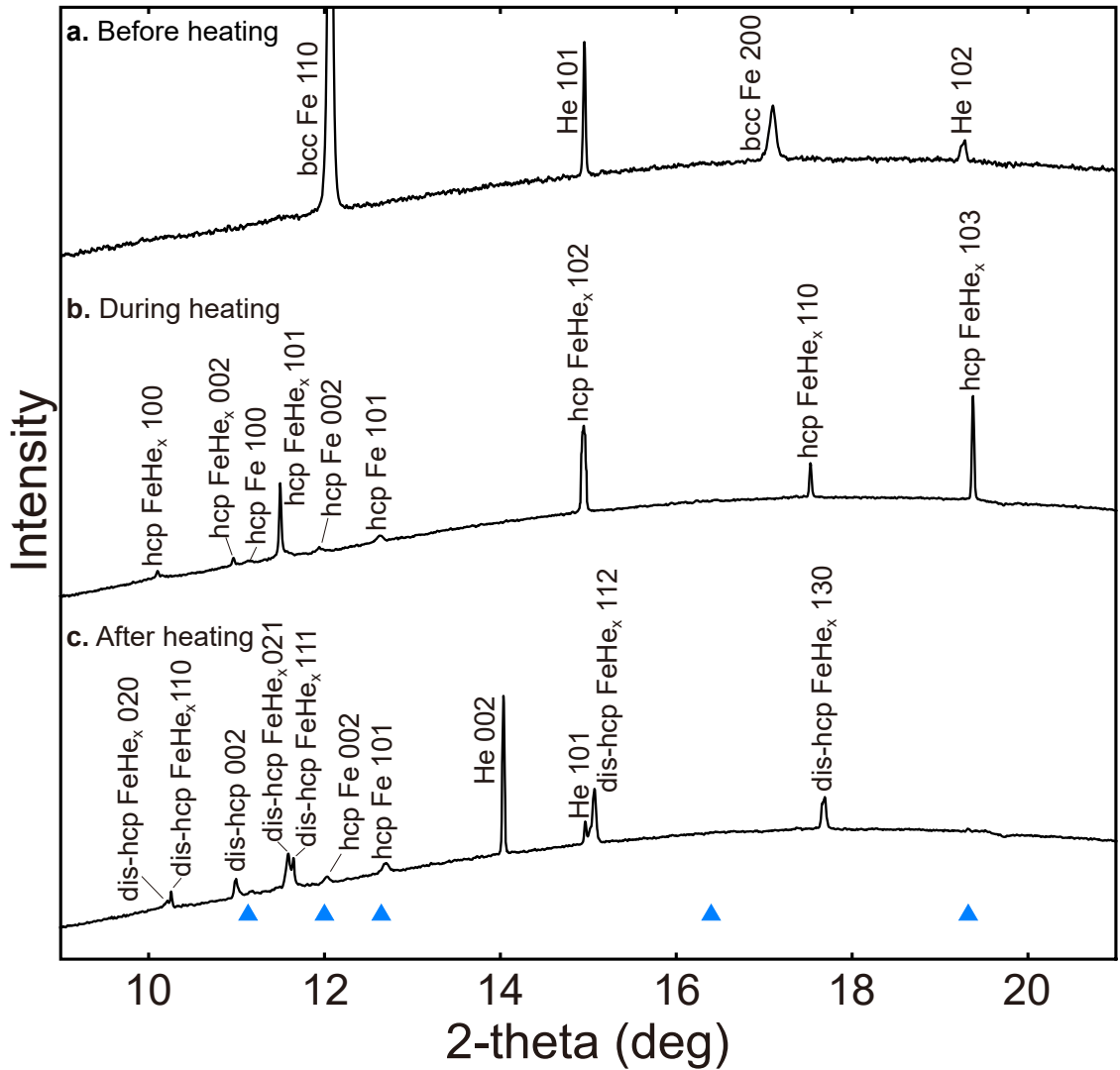

FIG. 1. XRD patterns collected in run #1 (a) before, (b) during, and (c) after heating. The peaks from hcp FeHe$_{0.48}$ appeared upon heating to 1450 K at 16 GPa, whose volume was substantially larger than that of pure Fe [see blue triangles in (c) indicating the peak positions for pure hcp Fe]. The minor peaks from hcp pure Fe were also included in (b) and (c). The hcp FeHe$_{0.48}$ exhibited crystallographic distortion to an orthorhombic structure (distorted hcp) upon quenching temperature.

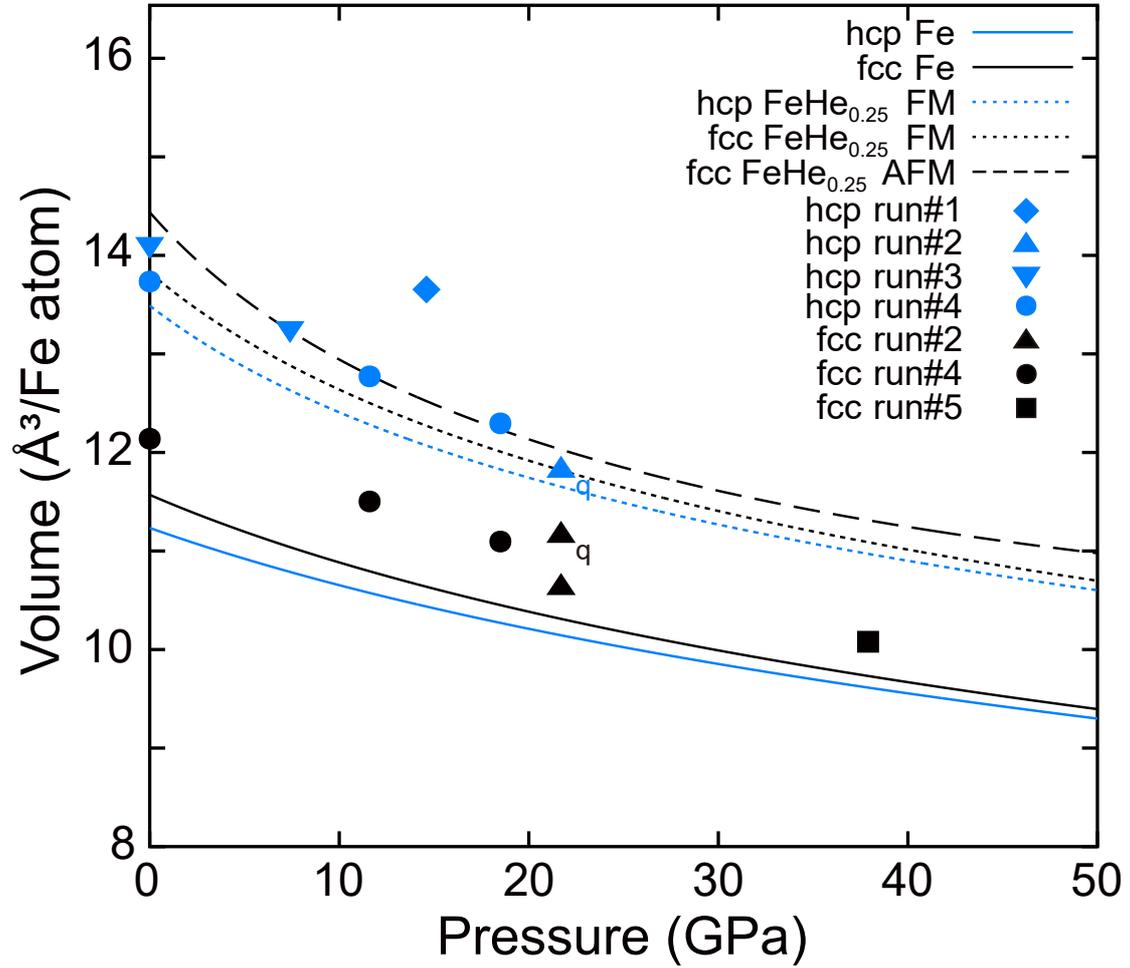

FIG. 2. The volumes per Fe atom of the expanded fcc and (distorted) hcp phases determined after quenching temperature to 300 K. Black, fcc; blue, (distorted) hcp. 'q' denotes quench crystals formed from liquid. Their volumes are compared to those of fcc and hcp pure Fe ($V_{Fe}$, solid lines) previously determined by Refs. [48] and [49], respectively. The compression curves of hcp and fcc $FeHe_{0.25}$ in the FM (dotted lines) and AFM-d states (broken lines) are obtained from $V_{Fe}$ by these earlier experiments and $\Delta V_{He}$ by the present static calculations of pure Fe and $FeHe_{0.25}$ (Fig. S2 [23]).

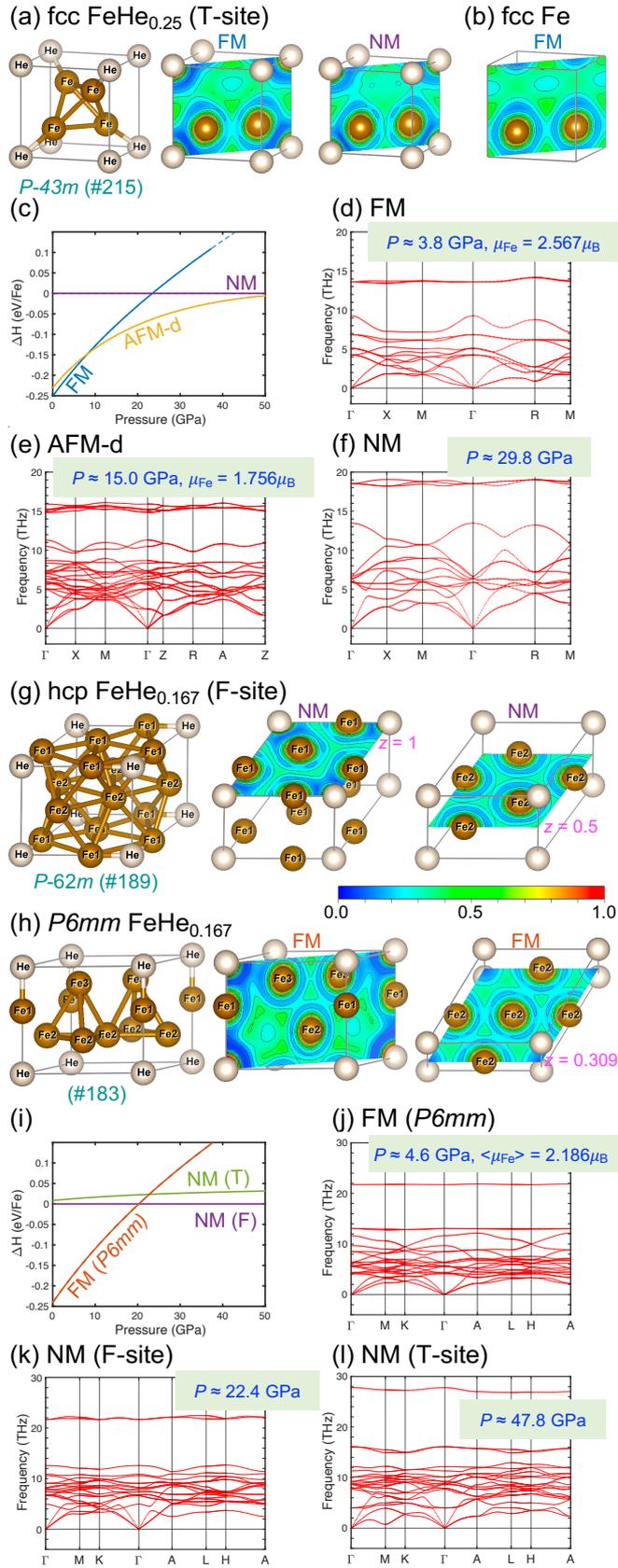

FIG. 3. Atomic structures, ELFs, phonon dispersions, and relative enthalpies ($\Delta H$) of fcc FeHe$_{0.25}$ (a–f) and hcp FeHe$_{0.167}$ (g–l) in various magnetic states and site occupancies. The nearest Fe-He distance is 1.843 Å in FM fcc FeHe$_{0.25}$ at 3.8 GPa (a) and 1.603 Å in NM hcp FeHe$_{0.167}$ at 22.4 GPa (g). Dashed lines in panel (c) indicate extrapolation; letters T and F in panel (i) indicate site occupancy of He. Banners in panels (d)–(f) and (j)–(l) indicate the static pressure ($P$) and Fe magnetic moment ($\mu_{Fe}$); <$\mu_{Fe}$> in panel (j) indicates the average magnetic moment of Fe1, Fe2, and Fe3 shown in panel (h). Detailed descriptions in the main text.

Supplemental Material for

# Formation of Iron-Helium Compounds under High Pressure

Haruki Takezawa,[1,*] Han Hsu,[2,†] Kei Hirose,[1,3,‡] Fumiya Sakai,[1] Suyu Fu,[1] Hitoshi Gomi,[3] Shiro Miwa,[4] and Naoya Sakamoto[4]

[1]*Department of Earth and Planetary Science, The University of Tokyo, Tokyo 113-0033, Japan*
[2]*Department of Physics, National Central University, Taoyuan City 320317, Taiwan*
[3]*Earth-Life Science Institute, Institute of Science Tokyo, Tokyo 152-8550, Japan*
[4]*Creative Research Institution (CRIS), Hokkaido University, Sapporo, Hokkaido 001-0021, Japan*

To whom correspondence should be addressed:
Haruki Takezawa (takezawa@eps.s.u-tokyo.ac.jp), Han Hsu (hanhsu@ncu.edu.tw), Kei Hirose (kei@eps.s.u-tokyo.ac.jp).

**High *P-T* experiments**

High *P-T* experiments were performed by using laser-heated DAC techniques. We used diamond anvils with 300 μm culet size. The culet was coated with a thin (<0.5 μm thick) layer of Ti by sputtering in order to prevent possible migration of helium into the diamonds [50]. For melting a sample in run #2, $Al_2O_3$ was additionally sputtered on one anvil for better thermal insulation. A rhenium gasket was preindented to 33–39 μm thickness. After a ~10 μm thick pure Fe foil (>99.999% purity, Toho Zinc) was put into a sample chamber, supercritical helium was loaded under ~200 MPa by using a high-pressure gas apparatus (PRETECH Co., Ltd) at SPring-8. A high-pressure vessel and a gas line were vacuumed for 10 min and then purged by helium gas four times before loading it into the DAC. Only in run #4, the <10 μm thick $Al_2O_3$ plates were employed for better thermal insulation, while large space was left for helium. The XRD patterns collected before and after heating at high pressures showed the peaks from hcp solid helium in all runs except run #3 in which helium was in a liquid state (Figs. 1, S3, S8, and S9).

After compression to a pressure of interest at room temperature, the iron foil was heated from both sides with a couple of 100 W single-mode Yb fiber lasers at the beamline

BL10XU, SPring-8. A laser beam was converted to one with a flat energy distribution by beam-shaping optics, and the laser-heated spot was ~20–30 μm across. The one-dimensional temperature distribution across a laser-heated spot was obtained by a spectro-radiometric method [51]. Synchrotron XRD patterns of the sample were collected on a flat panel X-ray detector (Varex Imaging, XRD1611CP3) before/during/after laser heating under high pressure. The incident X-ray beam was monochromatized to a wavelength of 0.41298–0.41530 Å (~30 keV) and focused to 6 μm in diameter. The sample temperature is the average of those over 6 μm area at a hot spot. Pressure at 300 K was primarily determined from the unit-cell volume of solid helium [52] (Table I), which is consistent with those estimated based on the Raman shift of a diamond anvil [53] and from the unit-cell volume of $Al_2O_3$ and its equation of state [54] (Table SI). Since the XRD peaks of helium were not observed in run #3 and during decompression in run #4, we employed the pressure based on the volumes of bcc Fe [55] found in an unheated portion and $Al_2O_3$ respectively in these runs. Pressure at high temperature was obtained from the thermal equations of state of pure iron [48,49], whose volume is estimated from observed $V_{FeHe_x}$ as $V_{Fe} = V_{FeHe_x} - x \times \Delta V_{He}$ assuming that helium concentration $x$ and $\Delta V_{He}$ did not change upon quenching temperature. Indeed, pressure during heating at 1660 K in run #4 is estimated separately from the hcp and fcc phases to be 24.1 and 23.7 GPa, respectively, indicating the validity of this method.

We estimated the amounts of helium $x$ in $FeHe_x$ with expanded volumes by considering that helium, similar to hydrogen, is incorporated interstitially around Fe atoms (see the theoretical calculations section in the main text). The helium content $x$ was calculated from its lattice volume observed at high pressure and 300 K, assuming the lattice volume expands linearly with helium concentration;

$$x = \frac{V_{FeHe_x} - V_{Fe}}{\Delta V_{He}} \qquad (1)$$

The reference $V_{Fe}$ is from Refs. [48] and [49] for the fcc and hcp phases, respectively. The $\Delta V_{He}$ at high pressure is determined from the difference between the volumes of pure Fe and $FeHe_{0.25}$, which were obtained by static first-principles calculations in this study [Figs. S12(e) and Fig. S13(b), see Table SIV for parameters in the 3rd-order Birch-Murnaghan equations of state]; $\Delta V_{He} = 4 \times (V_{FeHe_{0.25}} - V_{Fe})$. Such $\Delta V_{He}$ is comparable to the volume of solid He per He atom at high pressures greater than ~5 GPa, while the latter is significantly larger below ~1 GPa [52] (Fig. S2). We consider the AFM-d state above 8 GPa and the FM state at lower pressures including 1 bar when estimating helium concentrations in fcc [Fig. 3(c), Table I, and Table SI]. The hcp and distorted hcp phases formed below 22 GPa in the present experiments could be in the FM state [Fig. 3(i)]. The hydrogen content in FeH$_x$ has been estimated in a similar manner (e.g., [56,57]). We assume the $\Delta V_{He}$ for hcp and distorted hcp to be identical with that for fcc calculated here, which is supported by a previous study on $\Delta V_H$ (an increase in $V_{Fe}$ by incorporating an H atom) [32,58]. The present estimates of $x$ in (distorted) hcp FeHe$_x$ may represent the upper bounds since $V_{Fe}$ used for the FM hcp phase may be larger than that provided by the experimentally derived equation of state of hcp Fe [49] due to different magnetic states.

**Experimental results of runs #4 and #5**

In run #4, the Fe + He sample was sandwiched between the Al$_2$O$_3$ thermal insulation layers. We melted a part of the sample by heating to >2300 K at ~30 GPa and found both fcc and distorted hcp at 300 K in an area ~10 μm away from the melted portion (Fig. S8). The XRD pattern did not change after we thermally annealed this portion at 1660 K for 5 min, demonstrating fcc FeHe$_{0.09}$ in the AFM state and distorted hcp FeHe$_{0.33}$ in the FM state. The XRD patterns obtained from different areas of the sample showed that these two phases were widely distributed with identical unit-cell volumes. Upon decompression to 1 bar, the bcc phase appeared without volume expansion, while volume-expanded fcc and distorted hcp were still present (Fig. 2).

Additionally in run #5 conducted at 54 GPa, fcc FeHe$_{0.06}$ in the AFM state was formed upon heating up to 2140 K on one side and 1670 K on the other side (Fig. S9), which likely caused melting in the Fe-He system locally near the sample surface. The XRD pattern collected after quenching temperature at 38 GPa demonstrated volume-expanded fcc FeHe$_{0.06}$ in the AFM state. We recovered this sample at ambient condition and examined for possible contamination by carbon (from diamond anvils), oxygen, rhenium (from gasket), and tungsten (from a needle used for loading an iron foil). The analyses of relatively large Fe grains that should have been heated to high temperatures (Fig. S10) showed <0.42 wt% C, <0.28 wt% O, <0.17 wt% Re, and <0.38 wt% W (Table SII). Such minimal contamination does not explain the large volume expansion by 3.6% for FeHe$_{0.06}$ compared to pure Fe.

**SIMS and EPMA analyses**

As a complement to XRD analyses, we examined the helium contents in Fe-He compounds by secondary ion mass spectrometry (SIMS, *CAMECA* ims-7f) using the $^{16}O_2^+$ primary beam with energy of 10 keV in order to detect weak signals of helium. In earlier SIMS measurements of helium, Cs$^+$ primary ions were used to collect HeCs$^+$ as secondary ions. In contrast, here we employed $^{16}O_2^+$ as primary ions and collected $^4$He$^+$ to avoid the elevation of background; Cs$^+$ ion intensities are so high that the tail of the Cs$^+$ peak overlaps with the HeCs$^+$ peak. We first obtained the two-dimensional, secondary ion image of $^{56}$Fe$^+$ with relatively low current (37 nA) and a large raster size (Fig. S19), which covered the entire Fe sample including a portion outside a sample chamber. Locating the primary beam position at the center of a laser-heated portion where iron-helium compound was formed in a DAC, the analysis of $^4$He was then performed with the intense primary beam (500 nA). The analysis area was 64 μm in diameter, and the pressure during measurements was $8.0 \times 10^{-7}$ Pa.

The first SIMS measurements were made under non-cryogenic temperatures on the FeHe$_{\sim0.4}$ sample, which was synthesized at 55 GPa and 1800 K and then recovered at ambient condition. The depth profile showed high concentration of helium at the surface of the sample but practically no signals from the inside (black profile in Fig. S6). On the other hand, the second sample recovered from run #2 was analyzed under cryogenic temperatures below -150 °C. We obtained signals of helium over the entire depth of the maximum ~1.5 μm thick sample (blue profile in Fig. S6). By employing the helium-implanted Fe foil (Ion Technology Center, 180 keV, ion dose $1 \times 10^{16}$/cm$^2$) as a standard (Fig. S20) and considering the 10 μm × 14 μm size of the liquid pool in run #2 based on two-dimensional XRD analyses (He-poor fcc FeHe$_{0.05}$ outside the liquid pool included a small amount of helium and therefore its contribution was small), the SIMS data provided 1.0 wt% He in quenched liquid.

In addition, we also recovered a sample from run #5 and examined possible contamination to iron by carbon, oxygen, rhenium, and tungsten on its cross section (Fig. S10 and Table SII). A sample cross section at the center of a laser-heated portion was prepared parallel to the compression axis using a focused Ga ion beam (dual beam FIB; Thermoscientific, Helios 5 UC). The scanning ion microscope (SIM) image was obtained with an accelerating voltage of 30 kV. The Fe, C, O, Re, and W contents in the Fe sample were quantified by a field emission-type electron probe micro analyzer, FE-EPMA (JEOL, JXA-8530F) with an acceleration voltage of 12 kV and a beam current of 15 nA. We used LIF (Fe, W), LDE2H (C), LDE1 (O), and PETH (Re) as analyzing crystals. Carbon concentration was quantified based on a calibration line obtained by a carbon-free copper mesh, Fe-0.84 wt%C (JSS066-6, the Japan Iron and Steel Federation), and Fe$_3$C [59]. Fe, W, Al$_2$O$_3$, and Re were also used as standards.

**Computational methods**

First-principles DFT calculations were performed using the Quantum ESPRESSO (QE) codes [29] for fcc FeHe$_{0.25}$ and hcp FeHe$_{0.167}$. We use the Perdew-Burke-Ernzerhof (PBE)-type generalized gradient approximation (GGA) for the exchange-correlation functional [60] and the projector-augmented wave (PAW) pseudopotentials available on the QE website (energy cutoff of 90 Ry). The k-point meshes for 5-atom fcc FeHe$_{0.25}$ and 7-atom hcp FeHe$_{0.167}$ cells are 10×10×10 and 9×9×8, respectively; equivalent k-point meshes were used for cells in other sizes. The DFT results are fitted to the 3rd-order Birch–Murnaghan equation of state (3rd BM EoS) [see Figs. S12(d), S13(a), S15(a), and Table SIV]. Phonon calculations were performed using the Phonopy package, in which the finite-displacement method is implemented [30]. In our phonon calculations, we use 135 or 180-atom supercells for fcc FeHe$_{0.25}$ (depending on the magnetic states) and 189-atom supercells for hcp FeHe$_{0.167}$.

TABLE SI. Experimental results. The fcc phase is considered to be in the AFM and FM states above and below 8 GPa, respectively. Helium concentrations calculated for the other magnetic state are given in italics. The hcp phase may be in the FM below 22 GPa.

| | Sample (+ medium) | $P$ (GPa)[a] by He | by Raman[b] | $T$ (K) | Observed phases | fcc $V$ (Å³)/Fe atom | $\Delta V$ (Å³)[e] | He content, $x$ AFM | FM | hcp, distorted hcp $V$ (Å³)/Fe atom | $\Delta V$ (Å³)[e] | He content, $x$ FM |
|---|---|---|---|---|---|---|---|---|---|---|---|---|
| 1 | Fe + He | 15.5 | 19.7 | 1450 | hcp Fe-He, hcp Fe | | | | | | | |
| | | 14.6 | 19.7 | 300 | distorted hcp[c] FeHe$_{0.48}$, hcp Fe, solid He | | | | | 13.581(98) | 3.144 | 0.481 |
| 2 | Fe + He | 30.5 | 33.7 | 2820 | fcc Fe-He | | | | | | | |
| | | 21.7 | 25.6 | 300 | fcc(q)[d] FeHe$_{0.13}$, fcc FeHe$_{0.05}$, hcp(q)[d] FeHe$_{0.28}$, solid He | 11.184(4) | 0.873 | 0.127 | *0.145* | 11.851(11) | 1.707 | 0.283 |
| | | | | | | 10.665(20) | 0.355 | *0.052* | 0.059 | | | |
| 3 | Fe + He | n.d. | −5.3 | ~1000 | distorted hcp[c] Fe-He | | | | | | | |
| | | n.d. | 7.4 | 300 | distorted hcp[c] FeHe$_{0.33}$ | | | | | 13.249(76) | 2.461 | 0.334 |
| | | 0.0 | 0.0 | 300 | distorted hcp[c] FeHe$_{0.32}$ | | | | | 14.102(85) | 2.868 | 0.318 |
| 4 | Fe + He (+ Al$_2$O$_3$) | 23.7 | 25.5 | >2300/1660 | fcc Fe-He, distorted hcp[c] Fe-He | | | | | | | |
| | | 18.5 | 20.5 | 300 | fcc FeHe$_{0.09}$, distorted hcp[c] FeHe$_{0.33}$, solid He | 11.096(11) | 0.645 | 0.091 | *0.104* | 12.303(21) | 2.033 | 0.326 |
| | | n.d. | 11.6 | 300 | fcc FeHe$_{0.09}$, distorted hcp[c] FeHe$_{0.32}$ | 11.503(18) | 0.709 | 0.089 | *0.104* | 12.751(15) | 2.177 | 0.319 |
| | | 0.0 | 0.0 | 300 | fcc FeHe$_{0.06}$, distorted hcp[c] FeHe$_{0.28}$, bcc Fe | 12.140(5) | 0.570 | *0.050* | 0.063 | 13.727(5) | 2.493 | 0.276 |
| 5 | Fe + He | 53.5 | 56.3 | 2140 | fcc Fe-He, hcp Fe | | | | | | | |
| | | 37.9 | 40.6 | 300 | fcc FeHe$_{0.06}$, hcp Fe, solid He | 10.079(1) | 0.347 | 0.055 | *0.064* | | | |

[a]Thermal pressure was added for pressure at high temperature. [b]Pressures determined by the Raman shift of a diamond anvil except for runs #3 (bcc Fe) and #4 (Al$_2$O$_3$). [c]Orthorhombic structure with the axial ratio of b/√3a = 1.012 to 1.031 while 1.0 for hcp, showing the distortion in the a-b plane. [d]Quench crystals. [e]Volume difference per Fe atom between observed and pure Fe [48,49].

TABLE SII. FE-EPMA analyses of a sample recovered from run #5.

| Point #[a] | Fe | C | O | Re | W | Total |
|---|---|---|---|---|---|---|
| 1 | 99.76 | 0.10 | 0.13 | 0.07 | 0.24 | 100.29 |
| 2 | 99.51 | 0.21 | 0.19 | 0.01 | n.d. | 99.92 |
| 3 | 97.88 | 0.17 | 0.17 | 0.05 | n.d. | 98.27 |
| 4 | 98.42 | 0.32 | 0.21 | 0.12 | n.d. | 99.06 |
| 5 | 99.63 | 0.28 | 0.21 | 0.07 | n.d. | 100.20 |
| 6 | 98.55 | 0.42 | 0.28 | n.d. | 0.38 | 99.62 |
| 7 | 99.70 | 0.02 | 0.12 | 0.06 | 0.07 | 99.97 |
| 8 | 99.42 | 0.01 | 0.12 | 0.01 | n.d. | 99.57 |
| 9 | 99.07 | 0.18 | 0.13 | n.d. | n.d. | 99.39 |
| 10 | 98.25 | 0.26 | 0.20 | n.d. | n.d. | 98.70 |
| 11 | 98.34 | 0.11 | 0.17 | 0.04 | n.d. | 98.66 |
| 12 | 99.43 | 0.15 | 0.15 | n.d. | n.d. | 99.73 |
| 13 | 100.52 | 0.06 | 0.05 | n.d. | n.d. | 100.63 |
| 14 | 98.85 | 0.07 | 0.07 | 0.03 | n.d. | 99.01 |
| 15 | 97.39 | 0.09 | 0.04 | 0.07 | 0.21 | 97.79 |
| 16 | 98.52 | 0.07 | 0.07 | n.d. | n.d. | 98.66 |
| 17 | 99.51 | 0.07 | 0.11 | n.d. | n.d. | 99.69 |
| 18 | 100.18 | 0.08 | 0.09 | 0.08 | n.d. | 100.42 |
| 19 | 97.82 | 0.07 | 0.01 | n.d. | n.d. | 97.90 |
| 20 | 100.07 | 0.10 | 0.06 | 0.05 | n.d. | 100.28 |
| 21 | 99.27 | 0.10 | 0.12 | n.d. | n.d. | 99.49 |
| 22 | 99.05 | 0.11 | 0.03 | 0.17 | n.d. | 99.36 |
| 23 | 98.72 | 0.14 | 0.14 | 0.01 | 0.14 | 99.14 |
| 24 | 98.83 | 0.08 | 0.07 | n.d. | n.d. | 98.98 |

[a]See Fig. S10.

The C content tends to be higher for points analyzed later (the position # does not correspond to the sequence of analysis).

TABLE SIII. Lattice parameters for the distorted hcp phase.

| Run # | $P$ (GPa) | $T$ (K) | Helium contents | $V$ (Å$^3$)[b] | $a$ (Å) | $b$ (Å) | $c$ (Å) | $c/a$ | $b/(\sqrt{3})a$ |
|---|---|---|---|---|---|---|---|---|---|
| 1 | 14.6 | 300 | FeHe$_{0.48}$ | 13.581(98) | 2.674(5) | 4.688(6) | 4.333(6) | 1.620(4) | 1.012(2) |
| 2 | 21.7 | 300 | FeHe$_{0.28}$(q)[a] | 11.851(11) | 2.559(1) | 4.432(1) | 4.179(1) | 1.633(1) | 1.000(0) |
| 3 | 7.4 | 300 | FeHe$_{0.33}$ | 13.249(76) | 2.622(4) | 4.683(5) | 4.316(4) | 1.646(3) | 1.031(2) |
|   | 0.0 | 300 | FeHe$_{0.32}$ | 14.102(85) | 2.694(4) | 4.770(6) | 4.390(5) | 1.630(3) | 1.022(2) |
| 4 | 18.5 | 300 | FeHe$_{0.33}$ | 12.303(21) | 2.574(1) | 4.533(2) | 4.217(1) | 1.638(1) | 1.017(1) |
|   | 11.6 | 300 | FeHe$_{0.32}$ | 12.751(15) | 2.605(1) | 4.592(1) | 4.263(1) | 1.636(1) | 1.018(0) |
|   | 0.0 | 300 | FeHe$_{0.28}$ | 13.727(5) | 2.669(1) | 4.710(1) | 4.369(1) | 1.637(1) | 1.019(0) |

[a]Hcp (not distorted) quench crystals. The lattice parameters are calculated when considering the orthorhomic lattice for this hcp phase. $b/(\sqrt{3})a = 1$ for the hcp structure. [b]Volume per Fe atom.

TABLE SIV. Parameters in Birch-Murnaghan equations of state for fcc FeHe$_{0.25}$ and Fe.

|  | FeHe$_{0.25}$ (AFM) | FeHe$_{0.25}$ (FM) | FeHe$_{0.25}$ (NM) | Fe (AFM) | Fe (FM) | Fe (NM) |
|---|---|---|---|---|---|---|
| $V_0$ (Å$^3$) | 14.12 | 14.34 | 12.03 | 11.25 | 12.08 | 10.29 |
| $K_0$ (GPa) | 66.9 | 94.4 | 193.3 | 152.8 | 169.6 | 284.5 |
| $K'$ | 5.5 | 6.4 | 5.0 | 2.4 | 4.0 | 4.6 |

$P$ (GPa) = 1.5 $K_0(x^{-7} - x^{-5})[1 + 0.75(K' - 4)(x^{-2} - 1)]$ where $x = (V/V_0)^{1/3}$, and $K_0$ and $K'$ are isothermal bulk modulus and its pressure derivative, respectively, at ambient pressure.

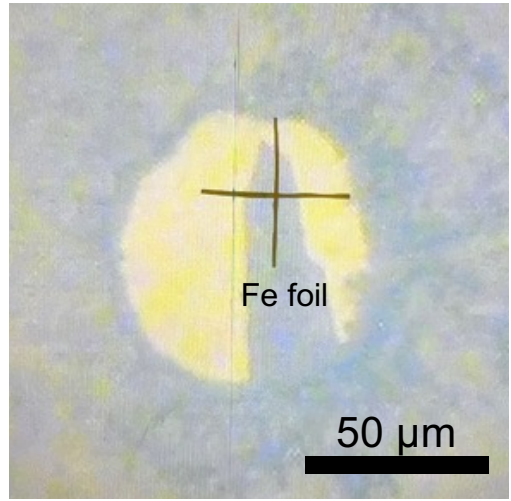

FIG. S1. Photomicrographs of a sample chamber after compression to 15 GPa in run #1. Plus (+) mark indicates the position of the center of a laser-heated spot.

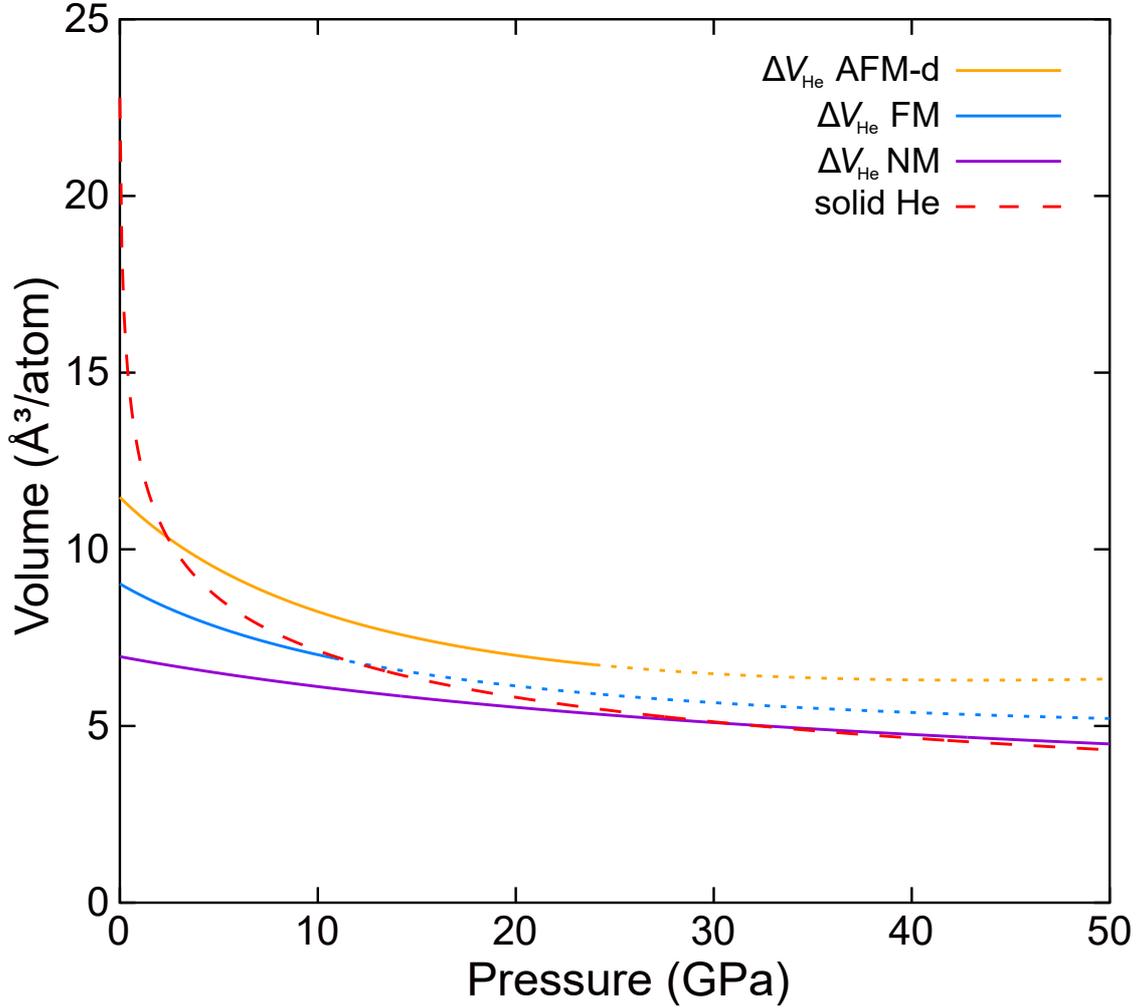

FIG. S2. $\Delta V_{He}$ (volume increase of Fe by incorporating a He atom) calculated as a difference between the volumes of Fe [Fig. S13(b)] and FeHe$_{0.25}$ from the present DFT calculations [Fig. S12(e)]; $\Delta V_{He} = 4 \times (V_{FeHe_{0.25}} - V_{Fe})$. Orange, blue, and purple curves represent $\Delta V_{He}$ in the AFM-d, FM, and NM states, respectively (dotted lines: extrapolation). The red broken line shows the volume of solid helium per atom reported by earlier experiments [52].

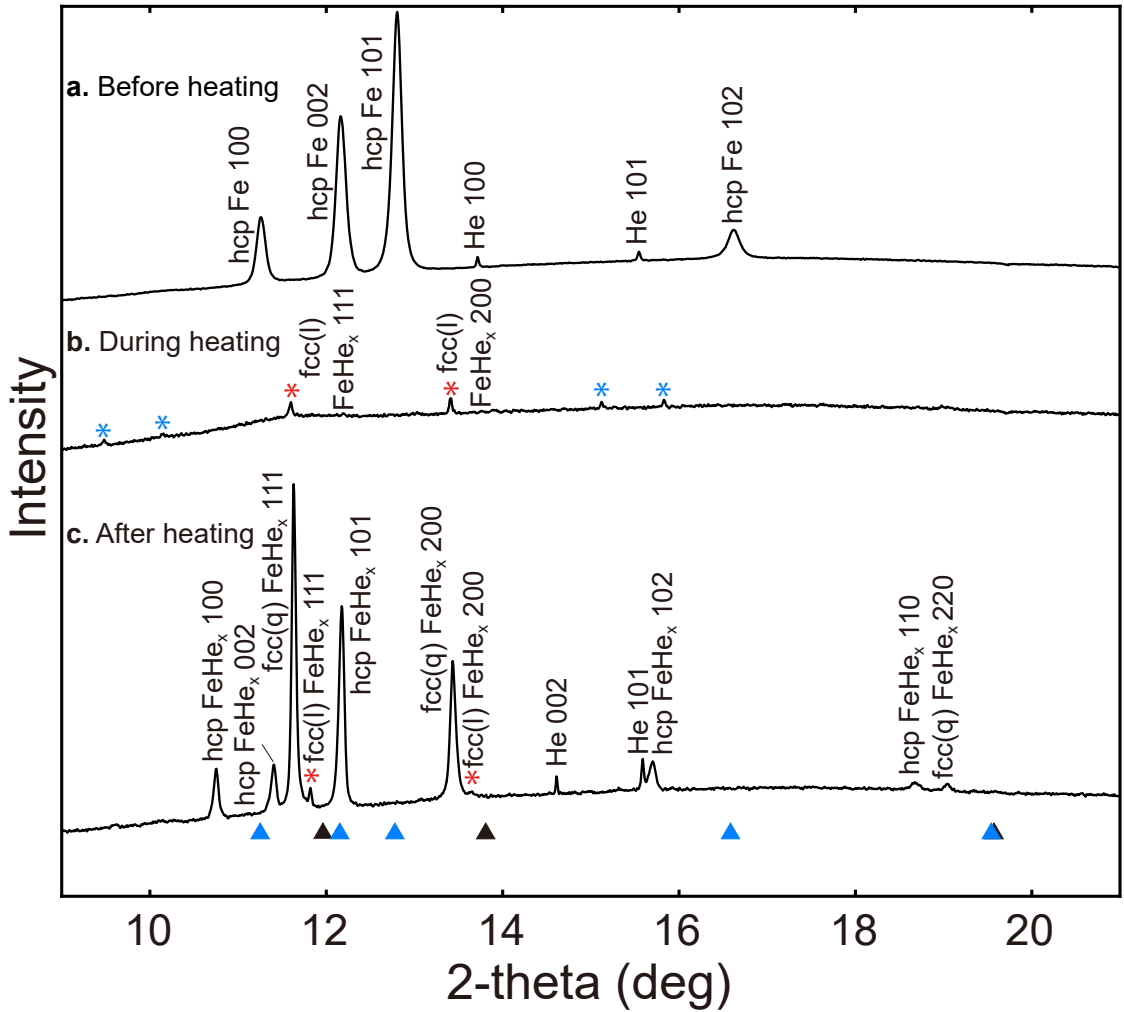

FIG. S3. XRD patterns collected in run #2 (a) before, (b) during heating to 2820 K at 31 GPa, and (c) after quenching temperature to 300 K. In (b), the diffuse scattering signal at 11–12 degrees of two theta angle shows the presence of liquid, coexisting with fcc (l) FeHe$_{0.05}$ (marked with red asterisks). Blue asterisks indicate minor peaks from Al$_2$O$_3$ sputtered onto diamond anvils. Upon quenching temperature, peaks from fcc FeHe$_{0.13}$ and distorted hcp FeHe$_{0.28}$ appeared (c), suggesting that liquid helium concentration was between them. Black and blue triangles indicate the peak positions for fcc and hcp pure Fe, respectively.

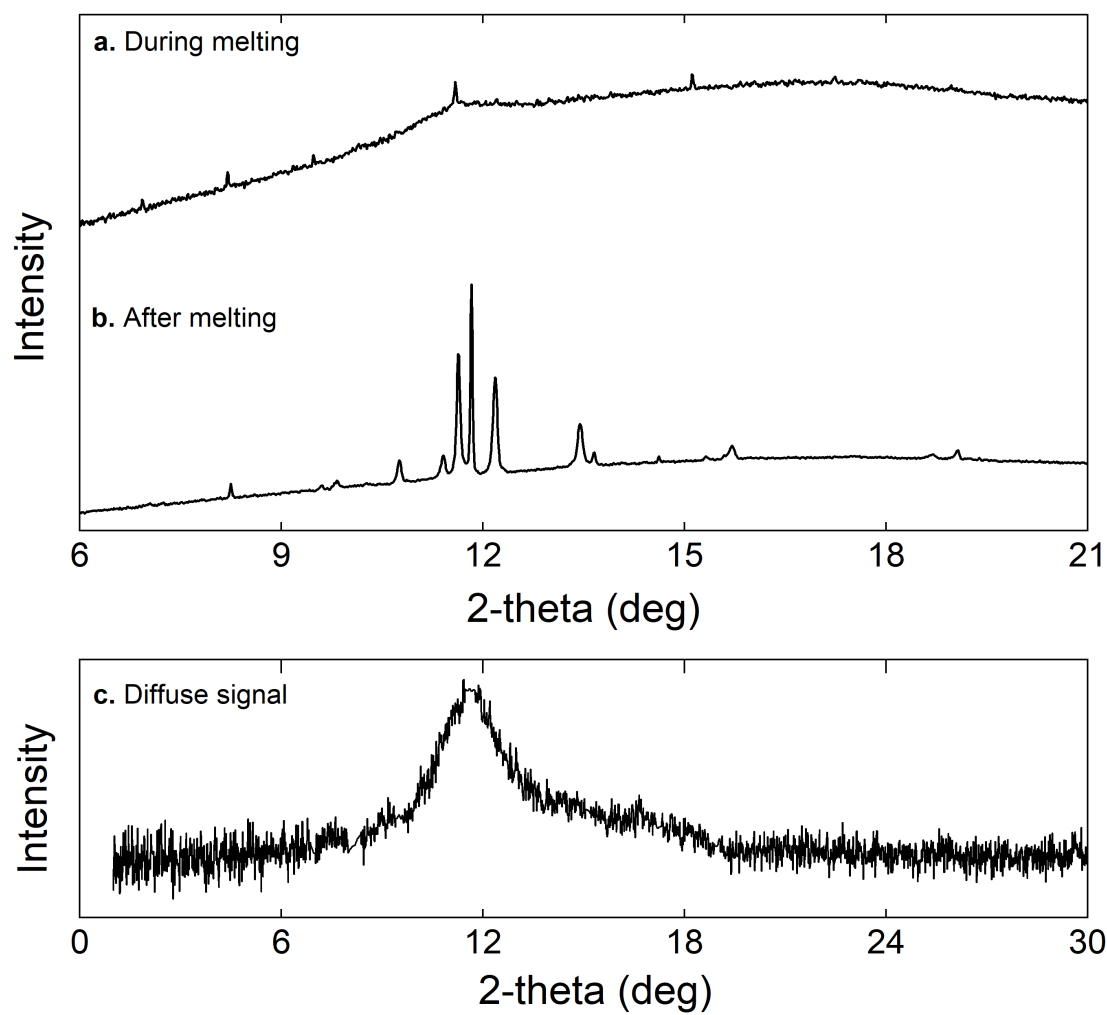

FIG. S4. XRD data in run #2 collected (a) during and (b) after melting. The diffuse signal from liquid (c) is obtained as the difference between these two patterns after removing peaks from solids.

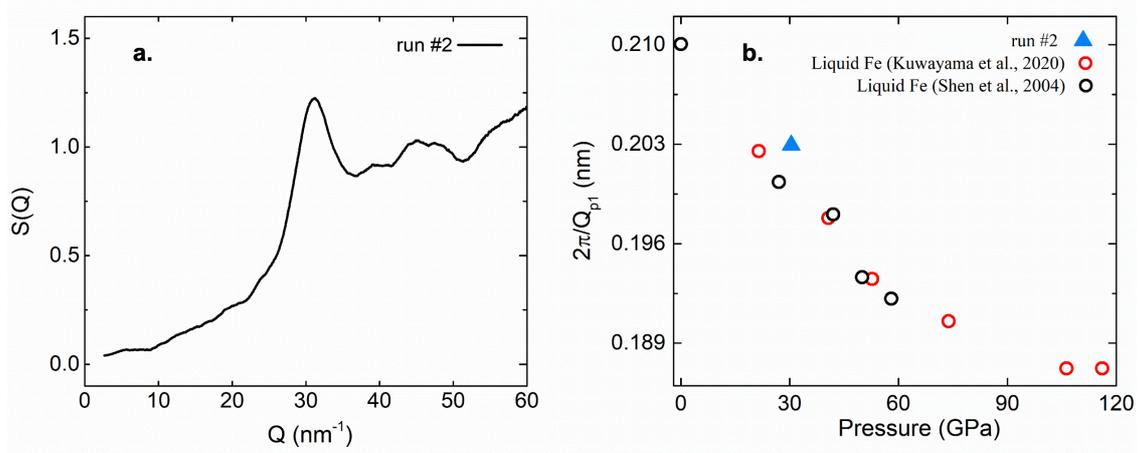

FIG. S5. (a) The structure factor S(Q) as a function of the momentum transfer Q, which was obtained from the diffuse signal observed in run #2 [see Fig. S4(c)]. (b) The $2\pi/Q$_distance derived from the first peak position in S(Q). The present datum for liquid FeHe$_{0.13-0.28}$ (blue triangle) is longer than those of liquid pure Fe reported by [27,28], indicating the elongation of the Fe-Fe distance in liquid by incorporating He.

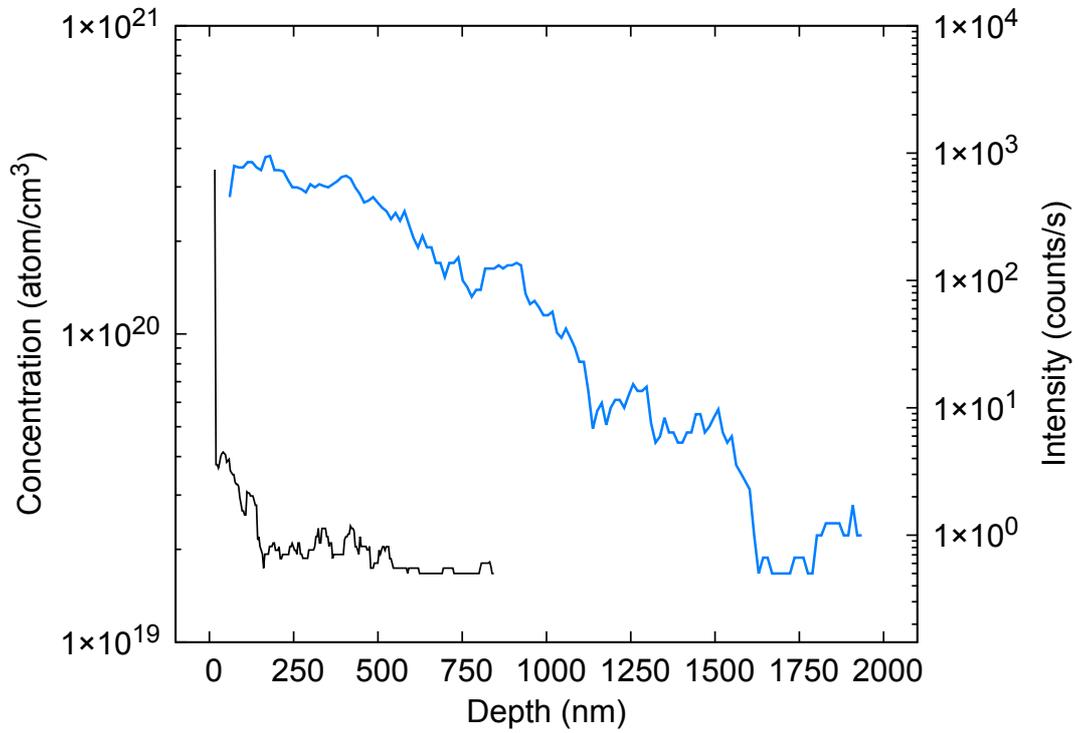

FIG. S6. Depth profiles of $^4$He$^+$ obtained by SIMS analyses of Fe-He samples under non-cryogenic (black line for intensity, see the right vertical axis) and cryogenic conditions (blue line for concentration, see the left vertical axis). Strong signal of helium was found only at the surface of the sample when measurement was performed under non-cryogenic temperatures. In contrast, when measured at cryogenic conditions, helium was observed from the entire depth of the maximum ~1.5 μm thick sample recovered from run #2 (nonuniform thickness of a melted part of the sample caused gradual reduction in helium signals because a thinner portion was lost earlier by ion beam irradiation). Such remarkable contrast in the depth profile suggests that helium escaped from the Fe lattice during ion beam irradiation and accumulated at the sample surface under non-cryogenic temperatures.

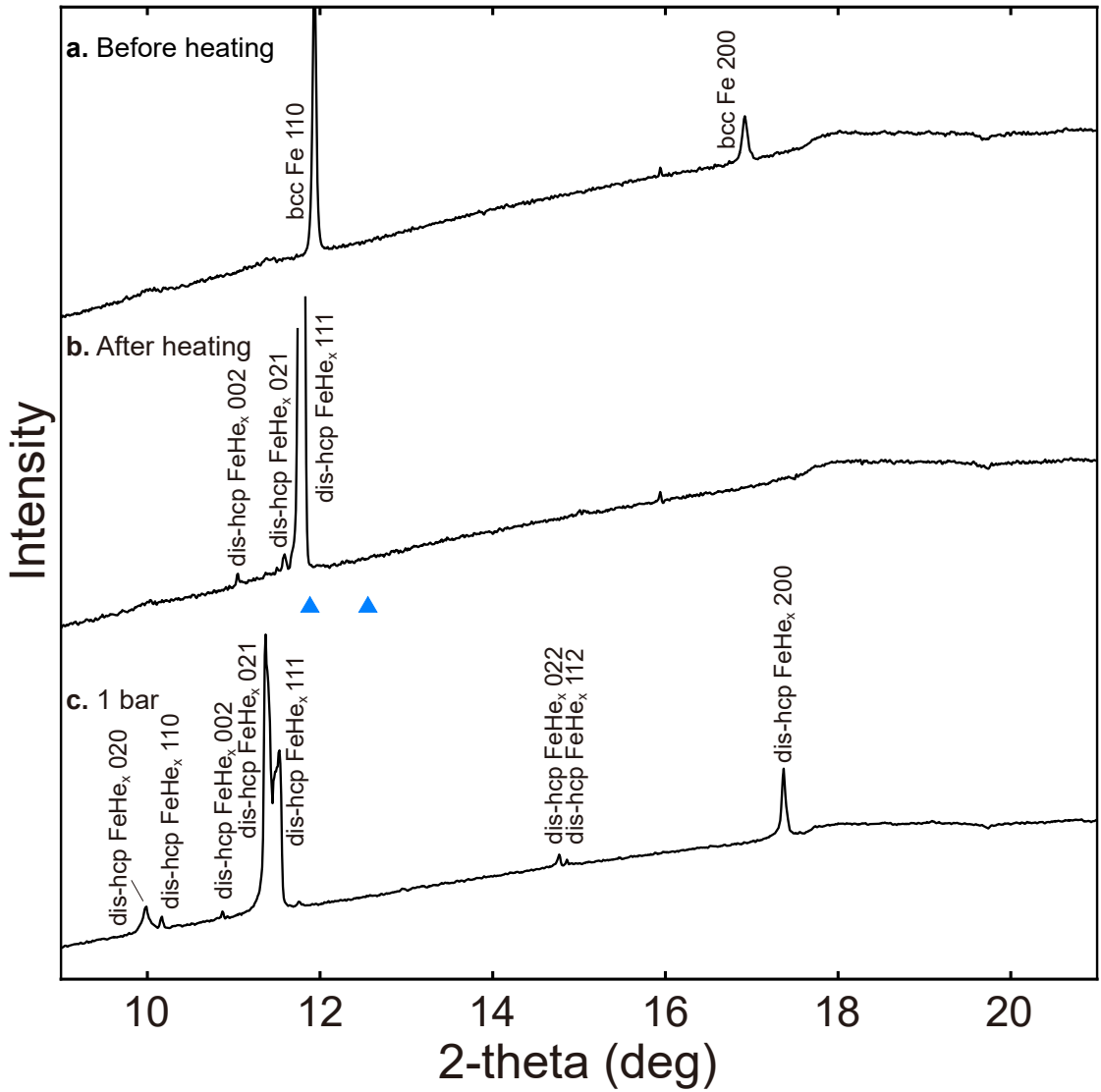

FIG. S7. XRD patterns collected in run #3 (a) before, (b) after heating at 7 GPa and 300 K, and (c) at 1 bar after decompression. The peaks from distorted hcp FeHe$_{0.33}$ appeared upon heating to ~1000 K at ~5 GPa, whose volume was substantially larger than that of pure Fe [see two blue triangles in (b) indicating the peak positions for pure hcp Fe 002 and 101]. This distorted hcp phase was still observed after complete pressure release, containing almost the same amount of helium.

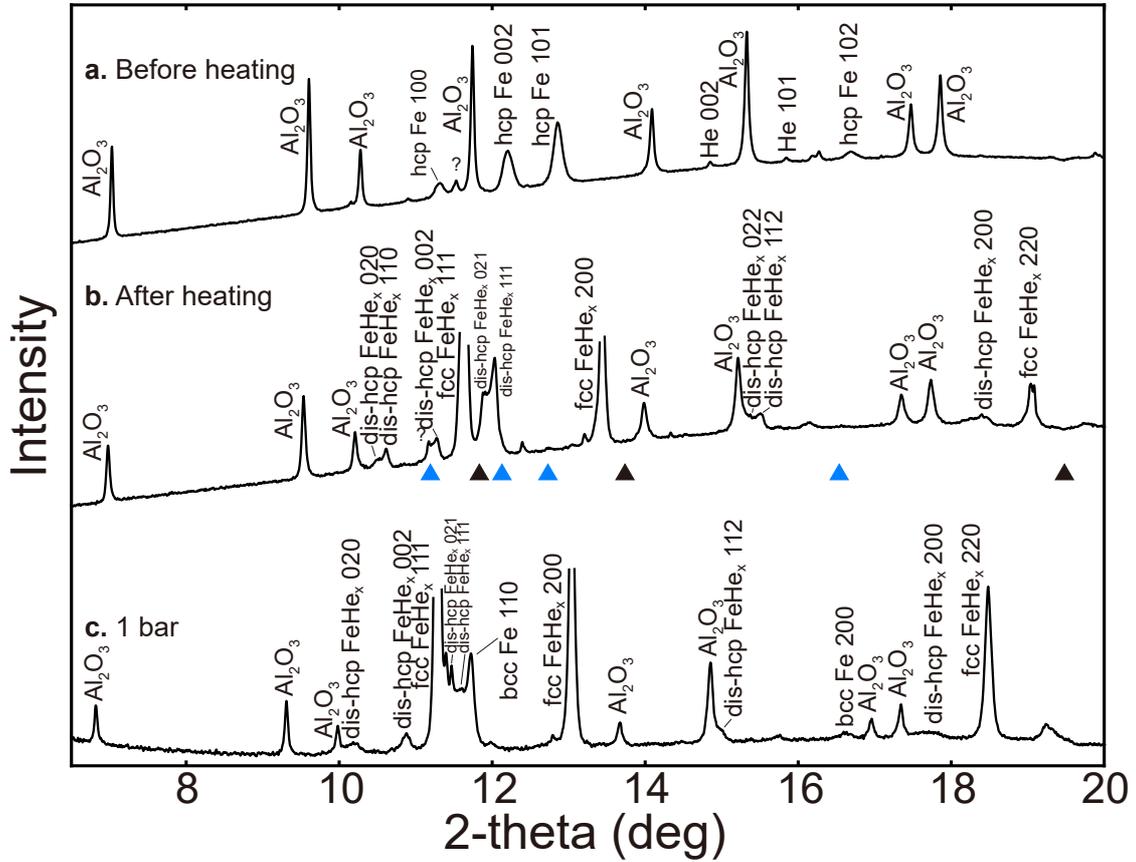

FIG. S8. XRD patterns collected in run #4 (a) before, (b) at 19 GPa after heating, and (c) upon decompression to 1 bar. The peaks from fcc FeHe$_{0.09}$ and distorted hcp FeHe$_{0.33}$ were observed after the sample was once heated to >2300 K. Their volumes were substantially expanded with respect to those of corresponding fcc and hcp pure Fe [see black and blue triangles in (b) indicating the peak positions for fcc and hcp pure Fe, respectively]. Upon decompression to 1 bar, bcc Fe appeared, while both fcc and distorted hcp were preserved. The helium content in the fcc phase reduced from $x = 0.09$ to 0.06, while that in distorted hcp did not change.

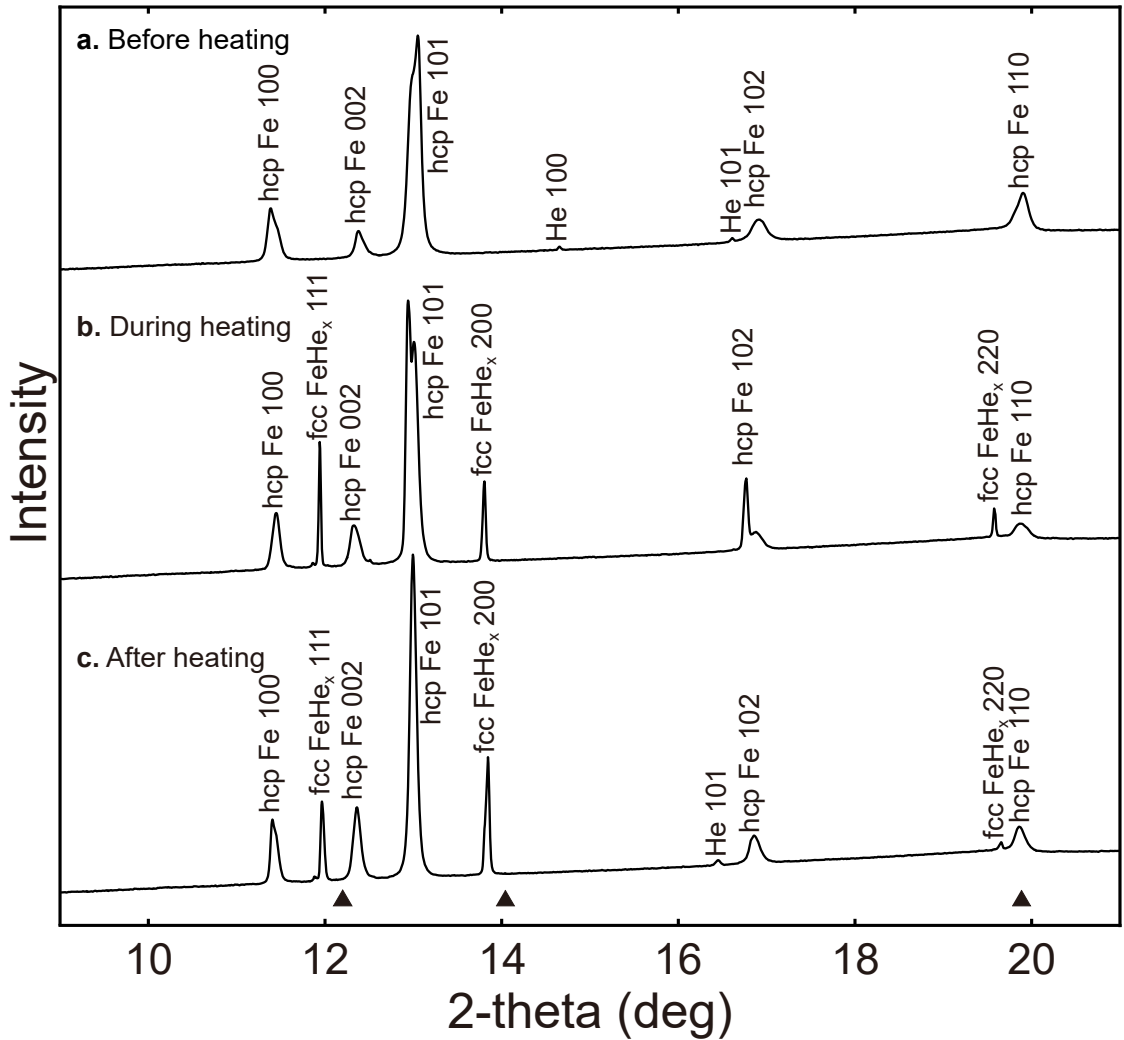

FIG. S9. XRD patterns collected in run #5 (a) before, (b) during, and (c) after heating. The peaks from fcc FeHe$_{0.06}$ appeared upon heating up to 2140 K at 54 GPa, whose volume was remarkably larger than that of pure Fe [see triangles in (c) indicating the peak positions for pure fcc Fe]. Because of a temperature gradient during laser heating and the tail of an X-ray beam, the XRD pattern during heating (b) included peaks from weakly heated and non-heated hcp crystals that were present somewhat away from the center of the hot spot (or in contact with a diamond anvil) and therefore did not undergo a phase transition to fcc.

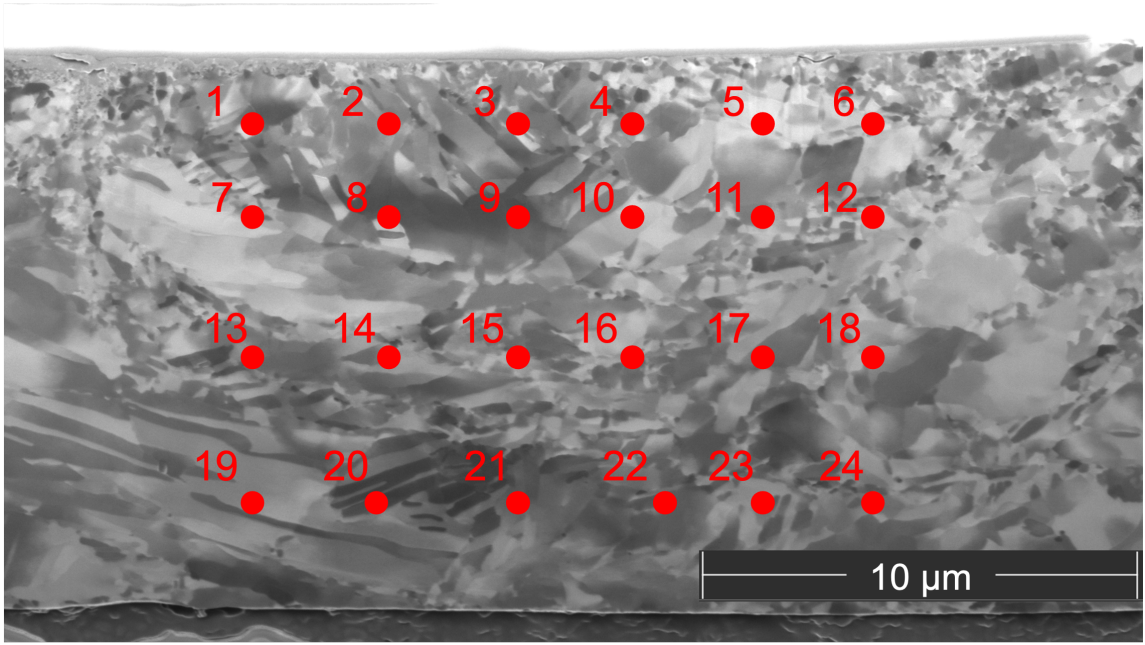

FIG. S10. Scanning ion microscope (SIM) image of a sample cross section at the center of the laser-heated hot spot for run #5. Red circles and numbers indicate the points of FE-EPMA analyses (see Table SII).

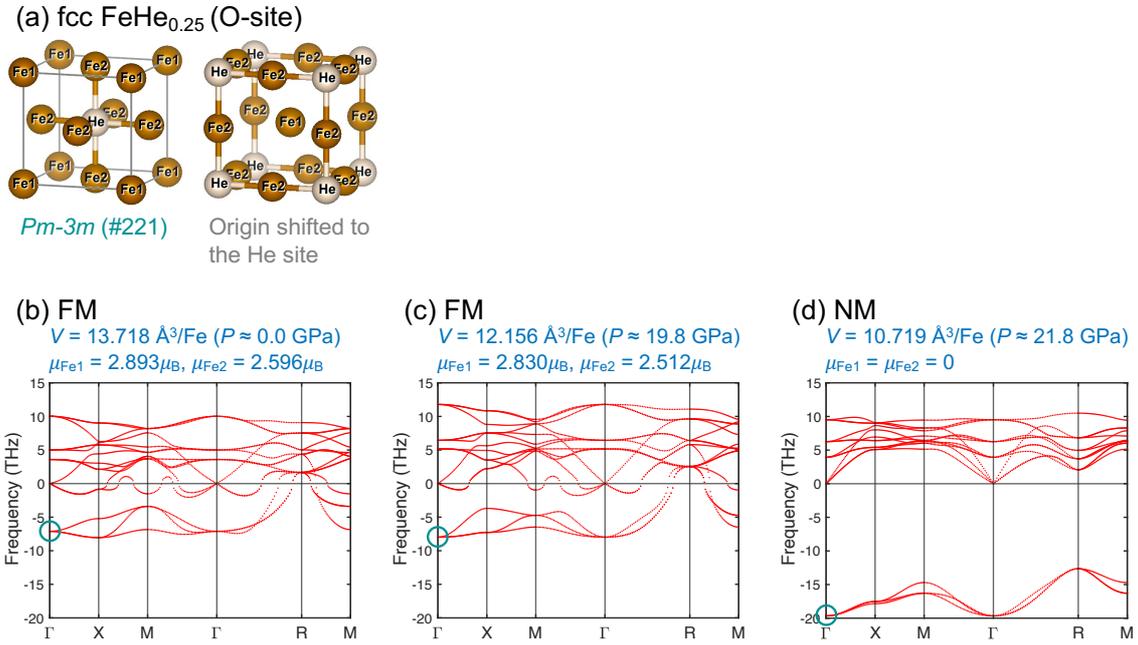

FIG. S11. (a) Atomic structure and (b)–(d) phonon dispersions of fcc FeHe$_{0.25}$ with O-site He. In this structure ($Pm$–$3m$), the two inequivalent Fe sites Fe1 and Fe2 are in the Wyckoff $1b$ and $3d$ positions respectively, with the He site placed at the origin ($1a$ position). This structure is dynamically unstable regardless the magnetic state or pressure, evident from the soft phonon modes shown in panels (b)–(d). At the Γ-point, the three degenerate soft phonon modes (indicated by circle) correspond to He displacements along the [100], [010], and [001] directions, respectively.

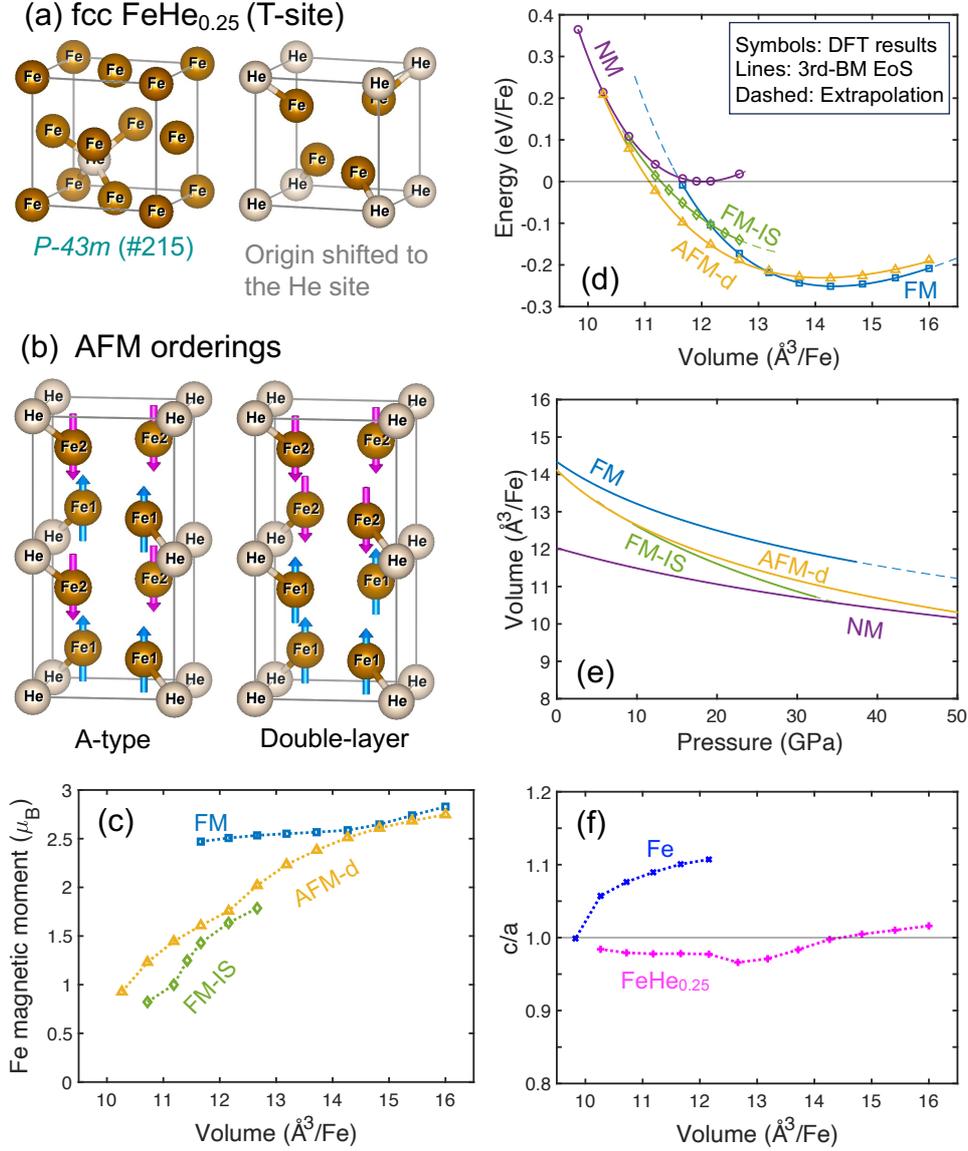

FIG. S12. (a) Atomic structure and (b) magnetic orderings of fcc FeHe$_{0.25}$ with T-site He. (c) Fe magnetic moment (as a functional of volume) of the obtained states, including two distinct ferromagnetically ordered states, referred to as the FM ($\mu_{Fe} \geq 2.5\ \mu_B$) and FM intermediate-spin (FM-IS, $\mu_{Fe} \approx 1$–$2\ \mu_B$) states. Symbols: DFT results; dotted lines: guide for the eye. (d) Equations of state $E(V)$ and (e) compression curves $V(P)$ of fcc FeHe$_{0.25}$ (T-site He). Dashed segments in panels (d) and (e) indicate extrapolation. (f) Tetragonal distortion ($c/a$) of AFM-d fcc Fe and FeHe$_{0.25}$ (T-site He); the latter is nearly cubic ($c/a \approx$ 0.97–1.00) in the region of 0–50 GPa ($14.118 < V < 10.309$ Å$^3$/Fe). The A-type AFM state has higher energy than the AFM-d state; it is therefore not included in the graphs.

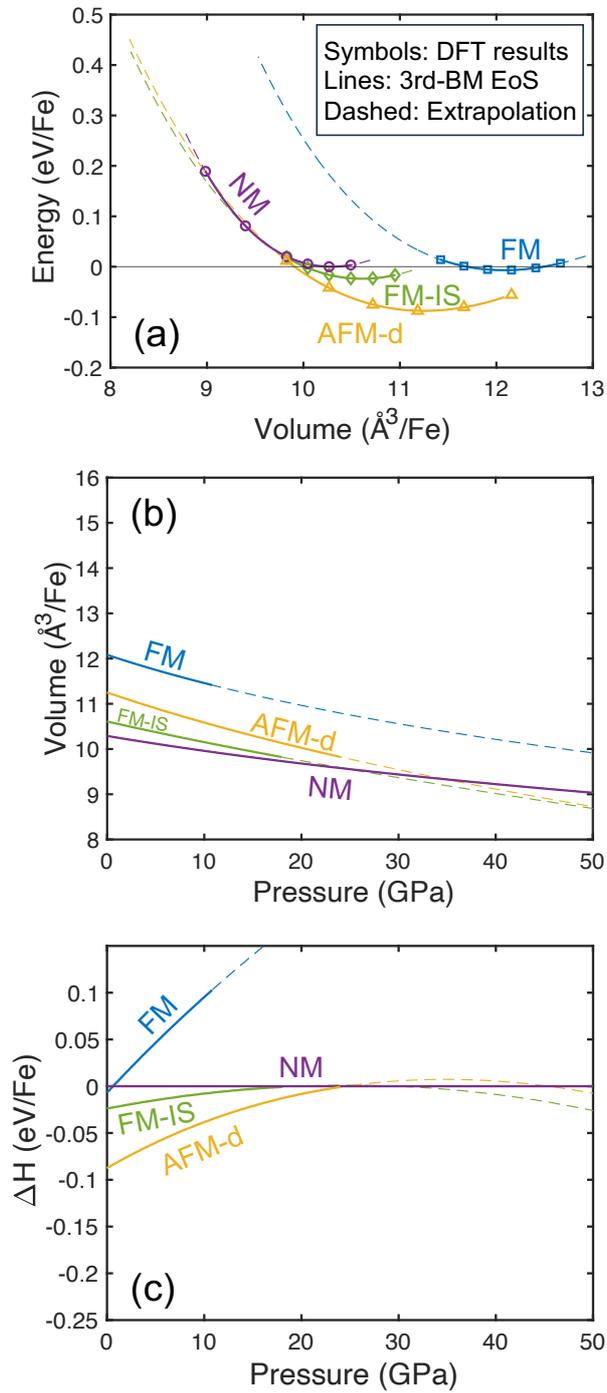

FIG. S13. (a) Equations of state $E(V)$, (b) compression curves $V(P)$, and (c) relative enthalpy $\Delta H$ (with respect to the NM state) of fcc Fe. Dashed segments indicate extrapolation.

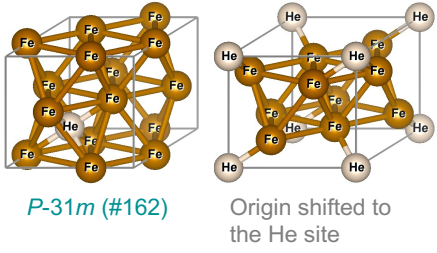
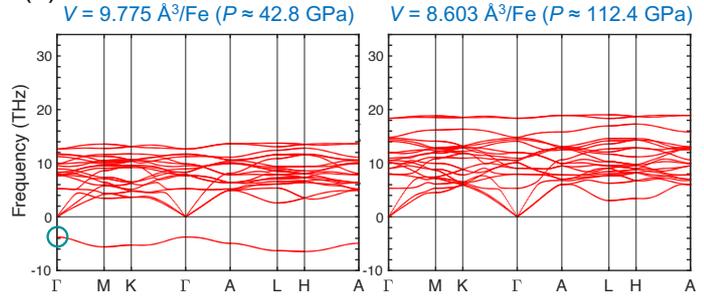
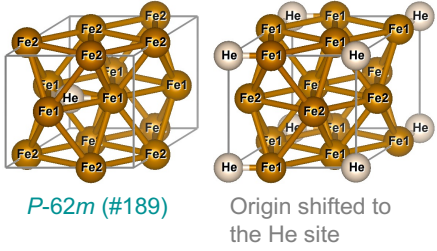
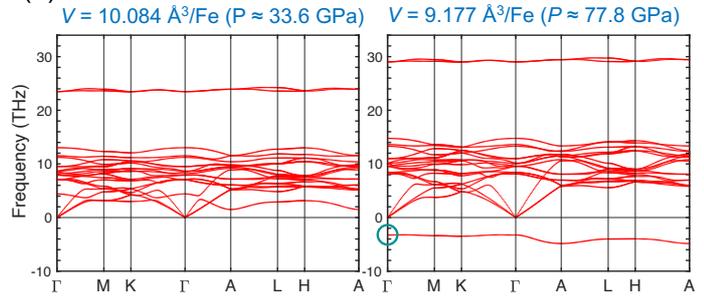
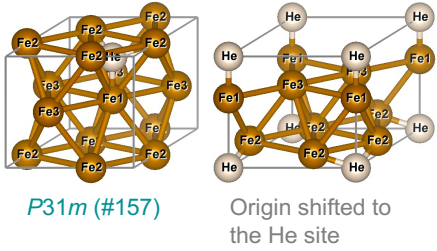
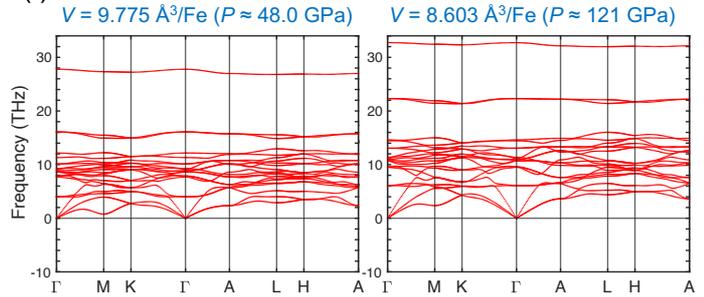

FIG. S14. Atomic structure and phonon dispersions of NM hcp FeHe$_{0.167}$ with (a, b) O-site, (c, d) F-site, and (e, f) T-site He. Circles in panels (b) and (d) indicate the Γ-point soft phonon modes; both modes correspond to He displacements along the [001] direction. In the $P31m$ structure [FeHe$_{0.167}$ with T-site He, panel (e)], the three inequivalent Fe sites Fe1, Fe2, and Fe3 are in the Wyckoff 1$a$, 3$c$, and 2$b$ positions, respectively.

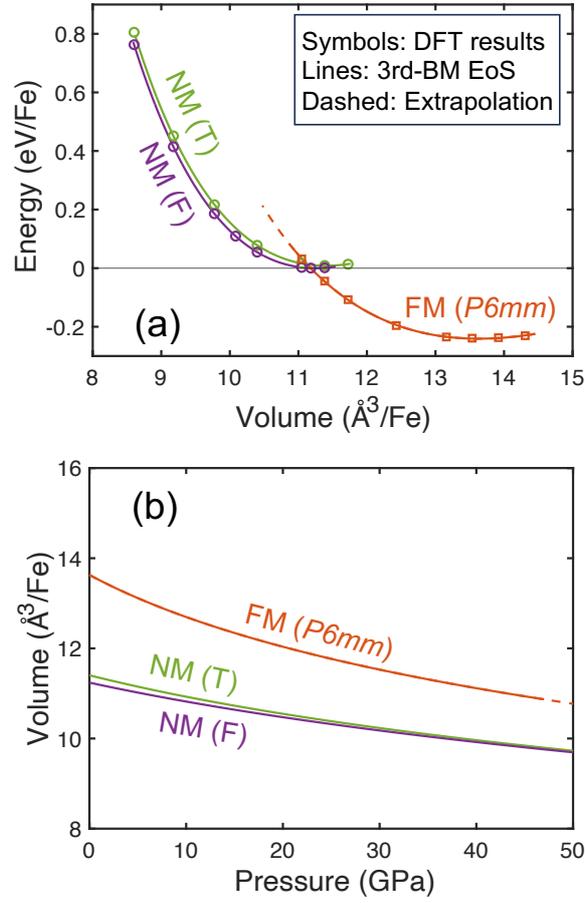

FIG. S15. (a) Equations of state $E(V)$ and (b) compression curves $V(P)$ of hcp FeHe$_{0.167}$. The dashed segments indicate extrapolation; letters T and F in panel (b) indicate site occupancy of He.

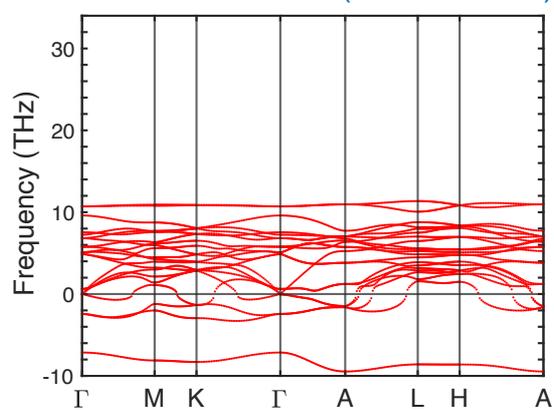

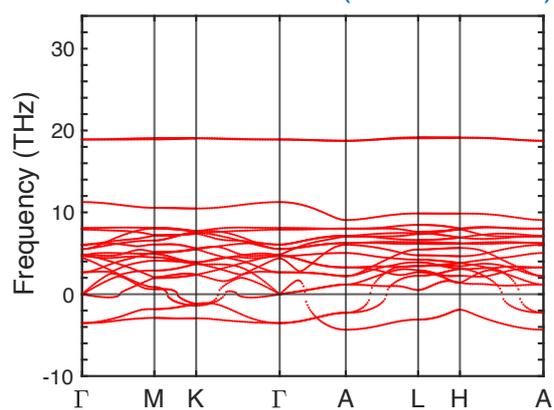

FIG. S16. Phonon dispersions of FM hcp FeHe$_{0.167}$ with (a) O-site and (b) F-site He at low pressure (< 20 GPa). Both structures are dynamically unstable.

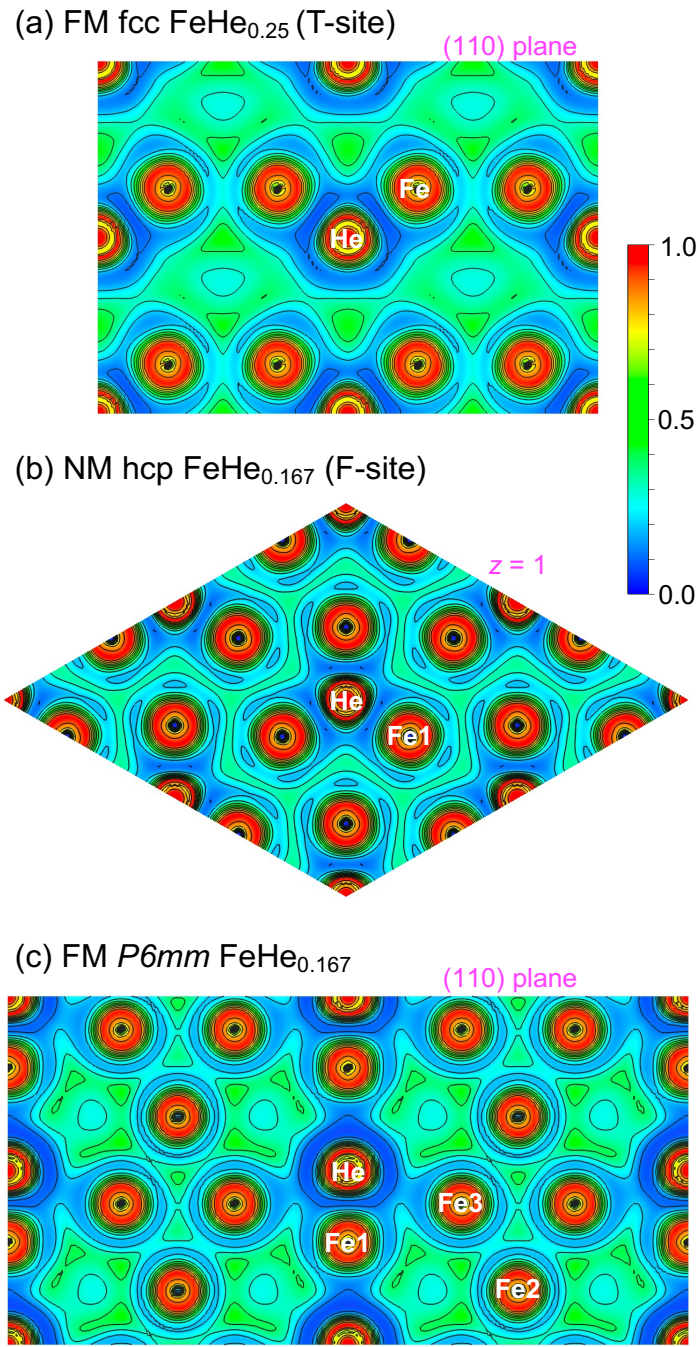

FIG. S17. The ELF of (a) FM fcc FeHe$_{0.25}$ (T-site He), (b) NM hcp FeHe$_{0.167}$ (F-site He), and (c) FM *P6mm* FeHe$_{0.167}$, corresponding to the ELF graphs in Figs. 3(a), 3(g), and 3(h), respectively, with more unit cells included. The He-Fe vdW interactions and the formation of Fe-He tetrahedra (a), trigonal planes (b), and dimers (c) can be observed.

(a) FM fcc FeHe$_{0.25}$ (T-site)   (b) FM fcc Fe

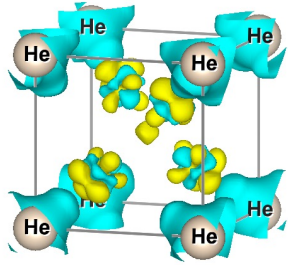 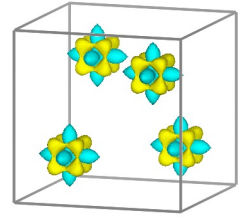

(c) NM hcp FeHe$_{0.167}$ (F-site)   (d) NM hcp Fe

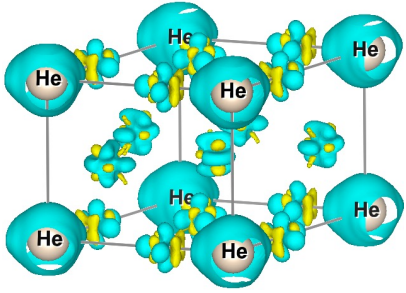 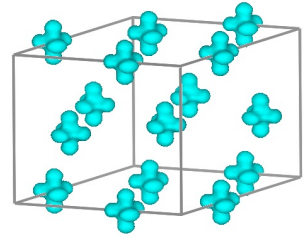

(e) FM *P6mm* FeHe$_{0.167}$

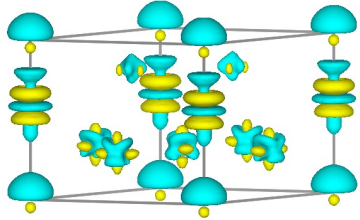

FIG. S18. Charge density difference ($\rho_{\text{diff}} \equiv \rho_{\text{crystal}} - \rho_{\text{atoms}}$, see main text) for fcc/hcp FeHe$_x$. Panels (a), (c), and (e) correspond to Figs. 3(a), 3(g), and 3(h) and also to Figs. S17(a)–S17(c). $\rho_{\text{diff}}$ for (b) FM fcc and (d) NM hcp Fe are also plotted for comparison. The isosurface values are $\pm 0.01$/bohr$^3$ (yellow and cyan, respectively). By comparing pure Fe [panels (b) and (d)] with FeHe$_x$ [panels (a) and (c)], electron charge transfer from He to Fe can be observed: negative lobes around the He sites reaching to the Fe sites, and positive lobes at the Fe sites facing the He sites. In FM *P6mm* FeHe$_{0.167}$, complicated dipole moments (in He and Fe1) are formed after the charge transfer from He to Fe.

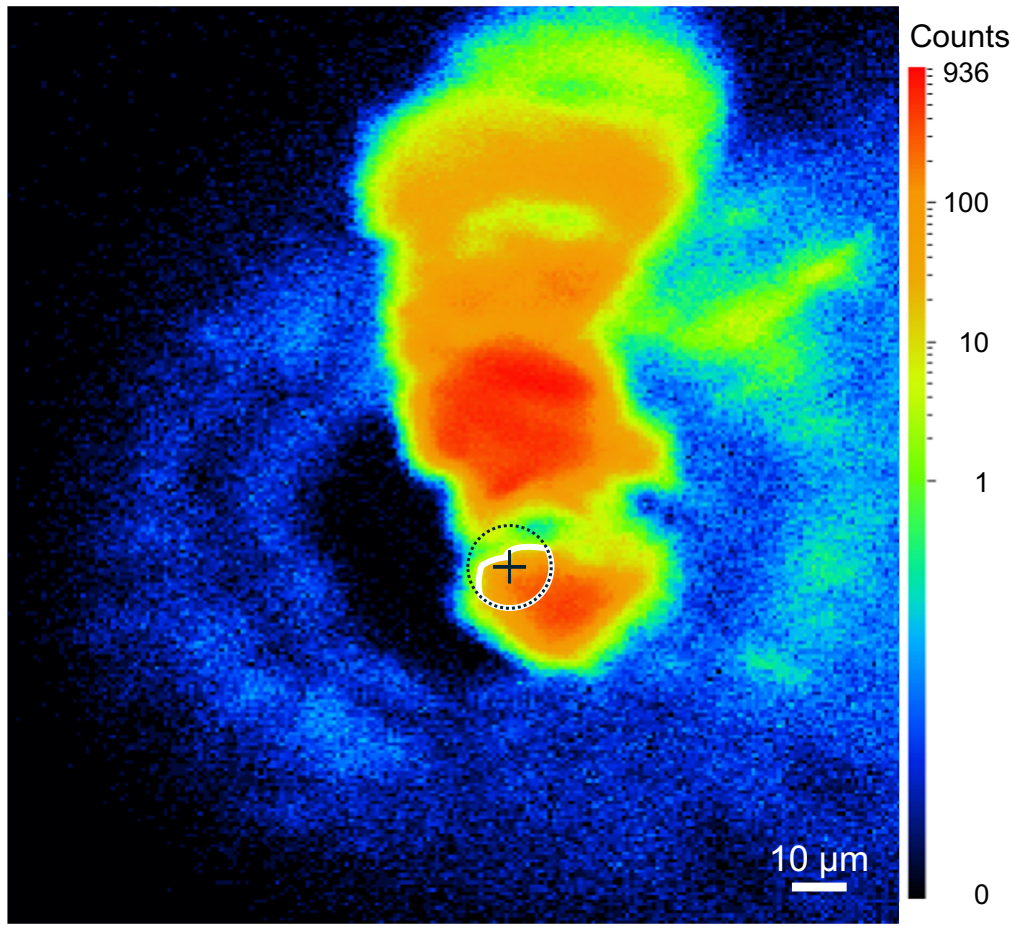

FIG. S19. Secondary ion image of $^{56}Fe^+$ showing the entire Fe sample including a portion outside a sample chamber, which was recovered from run #2. Plus indicates the position of the center of a laser-heated spot. The area enclosed by white line represents quenched liquid Fe-He after the sample shape slightly changed upon melting.

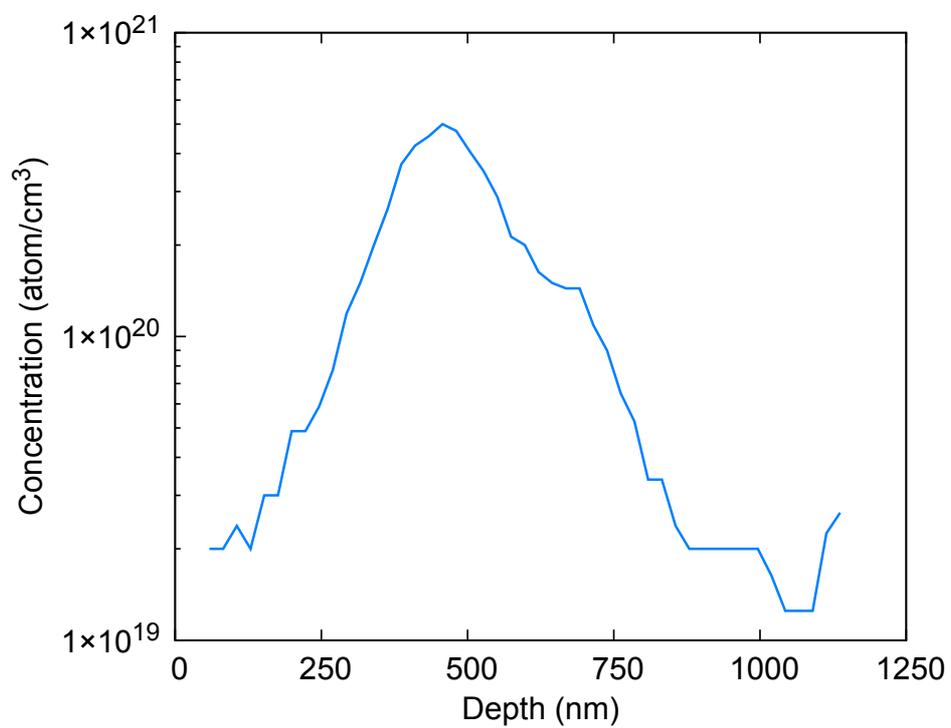

FIG. S20. SIMS analysis of helium in a helium-implanted Fe foil employed as a standard for the determination of He abundance in Fe-He compound recovered from run #2.